\documentclass[a4paper,11pt]{article}

\usepackage{preamble}

\title{Lepton number violation at hadron colliders via pseudo-Dirac heavy neutral leptons}

\author[basel]{Stefan Antusch\orcmail{0000-0001-6120-9279}{stefan.antusch@unibas.ch}}
\author[ist]{Jan Hajer\orcmail{0000-0001-8083-9102}{jan.hajer@tecnico.ulisboa.pt}}
\author[basel,lisboa]{Bruno M.\ S.\ Oliveira\orcmail{0000-0003-3293-6607}{b.m.silva.oliveira@tecnico.ulisboa.pt}}

\affiliation[basel]{Departement Physik, Universität Basel, Klingelbergstrasse 82, CH-4056 Basel, Switzerland}
\affiliation[ist]{Departamento de Física, Instituto Superior Técnico (IST), Universidade de Lisboa, 1049-001 Lisboa, Portugal}
\affiliation[lisboa]{Centro de Física Teórica de Partículas (CFTP), Instituto Superior Técnico (IST), Universidade de Lisboa, 1049-001 Lisboa, Portugal}

\AtBeginDocument{
  \hypersetup{
    pdfsubject = {High Energy Physics},
    pdfkeywords = {Lepton Number Violation, BSM Phenomenology},
  }
}

\begin{document}

\maketitle

\begin{abstract}
  Symmetry-protected low-scale seesaw models can account for the observed neutrino flavour oscillations without fine-tuning, while yielding collider-accessible signatures through pseudo-Dirac \HNLs.
  Seesaw frameworks generically predict \LN violation, which provides a powerful discovery channel.
  In symmetry-protected realisations, however, the amplitudes for \LN violation are strongly suppressed by destructive interference between the contributions of the two quasi-degenerate \HNLs within the usual $\QFT$ plane-wave treatment.
  We demonstrate that damped \NNOslong significantly alleviate this suppression.
  We compare the sensitivities to pseudo-Dirac \HNLs in both \LN-blind and \LN-violating channels at the $\LHC$ and future hadron colliders such as the $\FCChh$ and the $\SppC$.
  We find that, although searches for \LN violation outperform their \LN-blind counterparts, small mass splittings in the pseudo-Dirac \HNL pair can drastically reduce the sensitivities in these channels.
  We further show that combining \LN-blind and \LN-violating searches can distinguish a pseudo-Dirac \HNL pair from the double-Majorana limit in the intermediate regime where \LN violation is observable but not yet saturated.
\end{abstract}

\clearpage

\tableofcontents

\listoftables

\listoffigures

\clearpage

\section{Introduction}

In the \SM of particle physics, neutrinos are strictly massless Weyl fermions.
This follows from the absence of any renormalisable operator capable of generating neutrino masses.
However, the experimental observation of neutrino flavour oscillations has firmly established that \LFs mix and, consequently, that neutrinos have mass \cite{Super-Kamiokande:1998kpq, Super-Kamiokande:1998uiq, SNO:2001kpb, SNO:2002tuh, KamLAND:2002uet, K2K:2004iot, MINOS:2006foh, T2K:2015sqm}.
This constitutes unambiguous evidence for physics beyond the \SM.
Explaining the smallness of neutrino masses while remaining consistent with the successful structure of the \SM is therefore one of the central challenges in contemporary particle physics.

A minimal extension of the \SM that allows for neutrino masses is obtained by introducing right-chiral neutrinos and hence Dirac mass terms.
In this case, reproducing the observed sub-\unit{eV} neutrino masses requires Yukawa couplings many orders of magnitude smaller than those of the charged fermions.
However, in an effective field theory approach, neutrino masses can be generated within the \SM by the dimension-five Weinberg operator \cite{Weinberg:1979sa}.
This operator violates total \LN by two units, intrinsically linking neutrino mass generation with \LN violation.

Seesaw mechanisms provide \UV completions of the Weinberg operator by introducing heavy \DOFs whose exchange violates \LN and generates neutrino masses at low energies.
In the conventional type~I seesaw, heavy sterile neutrinos with Majorana mass terms are added to the \SM, leading to light neutrino masses suppressed by the ratio of the \EW scale to the \NP scale \cite{Minkowski:1977sc, Yanagida:1979as, Gell-Mann:1979vob, Mohapatra:1979ia, Schechter:1980gr, Schechter:1981cv}.
Achieving the observed neutrino masses typically requires either very large Majorana masses or extremely small Yukawa couplings, rendering the resulting \HNLs unobservable.
Alternatively, neutrino masses may arise from delicate cancellations among parameters, a possibility that motivates more structured realisations of the seesaw framework.

Symmetry-protected type~I seesaw models, such as the inverse seesaw \cite{Nandi:1985uh, Mohapatra:1986bd, Mohapatra:1986aw} and the linear seesaw \cite{Akhmedov:1995ip, Akhmedov:1995vm, Malinsky:2005bi, Antusch:2017tud}, address this issue by invoking an approximate \LN-like symmetry.
In these scenarios, the light neutrino masses are proportional to small symmetry-breaking parameters.
At the same time, the heavy Majorana neutrinos are forced into quasi-degenerate pseudo-Dirac pairs, with their mass splitting governed by the same symmetry-breaking parameters.
The \SPSS \cite{Antusch:2015mia, Antusch:2016ejd} provides a unified and model-independent description of the collider-relevant dynamics of this class of constructions \cite{Antusch:2022ceb}.

For \HNLs lighter than a few tens of GeV, long lifetimes enable displaced-vertex searches; see \eg \cite{Bondarenko:2018ptm,Alimena:2019zri,Abdullahi:2022jlv,Abada:2022wvh} and also \cite{Antusch:2016vyf, Antusch:2017hhu, Drewes:2018gkc, Drewes:2019fou}, including searches in unconventional datasets such as heavy-ion collisions \cite{Drewes:2018xma, Drewes:2019vjy, ElFaham:2022vot}.
For heavier \HNLs, \LN violation at hadron colliders is a powerful discovery channel \cite{Antusch:2016ejd}.
\SS dilepton final states are particularly clean, with low \SM backgrounds, and are often regarded as smoking-gun signatures of Majorana neutrinos.
In symmetry-protected models, however, the pseudo-Dirac structure suppresses these signals: a standard plane-wave \QFT calculation gives \LN-violating rates proportional to the small symmetry-breaking parameter, making them too small to observe \cite{Kersten:2007vk}.

Crucially, the above argument does not take the \HNLs' finite lifetime and propagation into account \cite{Antusch:2020pnn, Antusch:2023nqd}.
In particular, the small mass splitting gives rise to \NNOs between the two \DOFs of the pseudo-Dirac pair \cite{Antusch:2017ebe}.
This phenomenon is closely analogous to neutral meson-antimeson oscillations.
For long-lived \HNLs, these oscillations allow measurable \LN violation \cite{Antusch:2022hhh, Antusch:2023nqd}.
At lepton colliders, \LN violation can be measured in final-state distributions \cite{Antusch:2023jsa, Antusch:2024otj, Antusch:2026xxx}.

At large masses, however, the oscillation length is typically much shorter than the relevant experimental length scales, and naive expectations suggest that \LN violation remains unobservable.
A consistent assessment requires a first-principles description of \HNL production, propagation, and decay that incorporates coherence and decoherence effects arising from finite lifetimes, spatial localisation, and detector resolution.
This can be achieved within the framework of \QFT with external wave packets, in which the external particles are treated as localised wave packets rather than plane waves \cite{Beuthe:2001rc, Antusch:2022ceb, Antusch:2023nqd}.

The inclusion of decoherence effects damps the oscillatory behaviour and introduces a critical threshold for the mass splitting \cite{Antusch:2023nqd}.
When damping dominates and the mass splitting exceeds this threshold, observable \LN violation can be significantly enhanced.
As we demonstrate below, this condition can be satisfied in a well-defined region of parameter space, rendering \LN violation effects observable at hadron colliders despite the approximate \LN-like symmetry.

The strong suppression of \LN-violating effects in \SPSSs makes their experimental exploration particularly challenging at the \LHC, where both the available \COM energy and the achievable luminosity limit the sensitivity to small mass splittings and rare \HNL signatures.
In contrast, future hadron colliders such as the \FCChh \cite{FCC:2018vvp} and the \SppC \cite{CEPC-SPPCStudyGroup:2015csa}, operating at \COM energies of order \qty{100}{TeV}, offer a qualitatively new opportunity to probe these scenarios.
The substantially enhanced production rates for \HNLs, combined with extended kinematic reach and improved sensitivity to \SS dilepton signatures, allow these machines to access regions of parameter space that remain entirely inaccessible at the \LHC.
In this context, the \LHC serves as a well-defined reference point, while the \FCChh and \SppC emerge as essential facilities for testing damping-enhanced \LN violation in the multi-\unit{TeV} regime accessible in the \SPSS.

The remainder of this paper is organised as follows.
In \cref{sec:SPSS}, we review the \SPSS, derive the \SPSS mass spectrum, and discuss how \NNOs and decoherence determine the relevant \LN-violation regimes.
In \cref{sec:processes}, we identify the production and decay channels used for \LN-blind and \LN-violating searches and discuss the existing \LHC bounds that must be reinterpreted.
In \cref{sec:event generation}, we describe the \MC event generation, signal simulation, scaling procedure, and background samples.
In \cref{sec:analysis strategy}, we present the cut-based preselection, \BDT analysis, statistical treatment, and criteria for reinterpreting \LHC bounds and distinguishing pseudo-Dirac \HNLs from the double-Majorana limit.
In \cref{sec:results}, we present the projected sensitivities and their interpretation in terms of the pseudo-Dirac mass splitting and inverse seesaw parameters.
Finally, we summarise our conclusions in \cref{sec:conclusion}.

\section{\sentence\SPSSlong} \label{sec:SPSS}

\resetacronym{SPSS}

In type I seesaw scenarios accessible at colliders, the smallness of the neutrino masses is ensured by cancellations among different \LN-violating contributions.
To avoid fine-tuning, seesaw models such as the linear and inverse seesaws employ an approximate \LN-like symmetry based on a group $G_L$.
This symmetry protection not only keeps the neutrinos light, but also forces pairs of heavy Majorana neutrinos into pseudo-Dirac particles.

For collider studies, only the most accessible pair of sterile neutrinos, $N_1$ and $N_2$, is relevant.
The \SPSS \cite{Antusch:2015mia, Antusch:2016ejd} therefore supplements the \SM Lagrangian with
\footnote{
  Although $m_M$ is a Dirac mass, in symmetry-protected seesaws it plays the role occupied by the Majorana mass in conventional type~I seesaws.
}
\begin{multline} \label{eq:LSPSS}
  \mathcal L_{\SPSS}
  = \widebar N_i i \slashed\partial N_i
  - \left(
  y_{1\alpha} \widebar{N_1^c} \widetilde H^\dagger L_\alpha
  + y_{2\alpha} \widebar{N_2^c} \widetilde H^\dagger L_\alpha
  \right. \\ \left.
  + m_M^{} \widebar{N_1^c} N_2
  + \frac{\mu_M^\prime}{2} \widebar{N_1^c} N_1
  + \frac{\mu_M^{}}{2} \widebar{N_2^c} N_2
  + \hc
  \right) + \dots ,
\end{multline}
where the ellipsis indicate possible terms with additional heavy neutrinos, which we assume to be negligible in collider studies.
Here, $N_i$ and $L_\alpha$ are left-handed spinors, $\widetilde H = i\sigma_2 H^*$, and Lorentz and gauge contractions are implicit.
After \EWSB, the Yukawa interactions with the \SM Higgs \VEV $v \approx \qty{174}{GeV}$ generate the Dirac masses
\begin{align} \label{eq:yukawa coupling}
  \vec m_D^{} & = v \vec y_1 , &
  \vec \mu_D^{} = v \vec y_2 .
\end{align}
The small breaking of the \LN-like symmetry, together with the charges given in \cref{tab:LNLS}, ensures the mass hierarchy
\begin{align} \label{eq:mass hierarchy}
  0   & \leq \mu \ll m ,                                      &
  \mu & \in \{\mu_M^{}, \mu_M^\prime, \abs{\vec \mu_D^{}}\} , &
  m   & \in \{\abs{\vec m_D^{}}, m_M^{}\} .
\end{align}
We do \emph{not} assume a seesaw hierarchy between the two \LN-conserving mass parameters.
Since the \LN-violating mass parameters break the \LN-like symmetry independently, they are not expected to be of the same order.
The mass matrix of the neutral fermions in the interaction basis $\vec n = (\vec \nu^\trans, N_1, N_2)^\trans$ reads
\footnote{
  We indicate matrices using bold font.
}
\begin{align} \label{eq:mass matrix}
  \mat M^n
   & = \mat M_L + \mat M_{\slashed L} ,                  &
  \mat M_L
   & = \begin{pmatrix}
         \mat 0          & \vec m_D^{} & \vec 0 \\
         \vec m_D^\trans & 0           & m_M^{} \\
         \vec 0^\trans   & m_M^{}      & 0
       \end{pmatrix} ,
   &
  \mat M_{\slashed L}
   & = \begin{pmatrix}
         \mat 0            & \vec 0       & \vec \mu_D^{} \\
         \vec 0^\trans     & \mu_M^\prime & 0             \\
         \vec \mu_D^\trans & 0            & \mu_M^{}
       \end{pmatrix} .
\end{align}

\begin{table}
  \begin{tabular}{*4c} \toprule
             & $L$ & $N_1$ & $N_2$ \\\midrule
    $U(1)_L$ & $1$ & $-1$  & $1$   \\\bottomrule
  \end{tabular}
  \caption[Charges of the lepton fields under the \LNlong-like symmetry]{
    Charges of the lepton fields under the \LN-like symmetry group, realised as $G_L = U(1)_L$.
  } \label{tab:LNLS}
\end{table}

\subsection{\sentence\LNlong-conserving Dirac limit} \label{sec:Dirac limit}

\begin{figure}
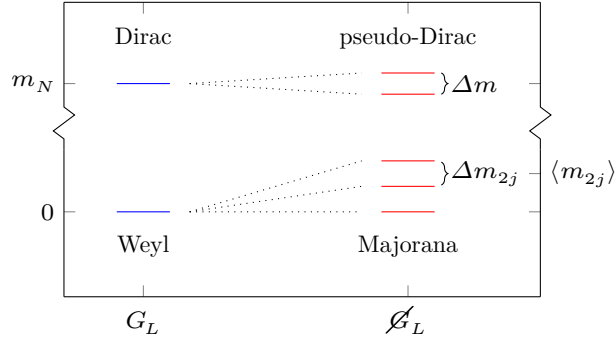

  \includepgf{spss-neutrino-mass-spectrum}
  \caption[Mass spectrum of the \SPSSlong]{
    Impact of the breaking of the \LN-like symmetry with group $G_L$ on the mass spectrum of the model.
    When the symmetry is unbroken, the three massless Weyl neutrinos of the \SM are accompanied by a single Dirac \HNL consisting of two degenerate Weyl \DOFs with mass $m_N$.
    A small breaking of the symmetry generates small Majorana mass terms for up to two of the \SM neutrinos, giving them an average mass of $\ev{m_{2j}}$ and a mass splitting of $\Delta m_{2j}$; additionally it splits the two heavy \DOFs by $\Delta m$.
  } \label{fig:mass splitting}
\end{figure}

The \LN-conserving part of the mass matrix \eqref{eq:mass matrix} has two degenerate singular values.
Hence, the five mass eigenstates $n_i$ fall into two classes: the three light neutrinos remain massless, while the two heavy \DOFs combine into a Dirac fermion with mass $m_N^{} = m_M^{} s$, where the singular value $s$ of the normalised mass matrix $\flatfrac{\mat M_L}{m_M}$ depends only on the active-sterile mixing vector
\begin{align} \label{eq:active-sterile mixing}
  s^2         & = 1 + \abs{\vec \theta}^2 ,        &
  \vec \theta & = \flatfrac{\vec m_D^{}}{m_M^{}} .
\end{align}
This spectrum is depicted in \cref{fig:mass splitting}.
As a consequence, in the \LN-conserving limit, the Lagrangian \eqref{eq:LSPSS} is identical to a model with a single Dirac \HNL.

The neutrino mixing matrix, which Takagi diagonalises the mass matrix, reads to all orders in the active-sterile mixing,
\begin{align} \label{eq:neutrino mixing matrix}
  \mat D_L & = \mat U_L^\trans \mat M_L \mat U_L , &
  \mat U_L & =
  \begin{pmatrix}
    \mat U_L^\nu                  & \frac{i\vec \theta^*}{\sqrt 2 s} & \frac{\vec \theta^*}{\sqrt 2 s} \\
    \vec 0^\trans                 & \frac{-i}{\sqrt 2}               & \frac{1}{\sqrt 2}               \\
    \frac{-\vec \theta^\trans}{s} & \frac{i}{\sqrt 2 s}              & \frac{1}{\sqrt 2 s}
  \end{pmatrix} .
\end{align}
The light neutrino mixing matrix
\footnote{
  Given two vectors $\vec u$ and $\vec v$, the matrix $\vec u \vec v^\trans = \vec u \otimes \vec v^\trans$ is their outer product.
}
\begin{equation}
  \mat U_L^\nu = \mat 1 - \frac{\vec \theta \vec \theta^\dagger}{s+s^2},
\end{equation}
is non-unitary
\footnote{
  Note that the light neutrino mixing matrix before \LN breaking is not the \PMNS matrix that describes the observed neutrino flavour oscillation data.
  After \LN breaking, the effective \PMNS matrix factorises into the product $U_0^\nu = U_L^\nu U_{\slashed L}^\nu$, where $U_{\slashed L}^\nu$ is its unitary component.
},
see \eg \cite{Antusch:2006vwa},
and its unitary defect is given by
\begin{equation}
  \mat \eta_L^\nu = (\mat U_L^\nu)^\dagger \mat U_L^\nu - \mat 1 = - \frac{\vec \theta \vec \theta^\dagger}{s^2} .
\end{equation}
Therefore, the norm
\begin{equation}
  \norm{\mat \eta_L^\nu} = \sqrt{\tr (\mat \eta_L^\nu)^\dagger \mat \eta_L^\nu} = \frac{\abs{\vec \theta}^2}{s^2} ,
\end{equation}
provides a scalar measure of how far the neutrino mixing matrix deviates from a unitary matrix.
It only becomes zero for vanishing active-sterile mixing parameters.

Apart from the \HNL mass, this model has three physical parameters, which can be taken to be the components of the real active-sterile mixing vector defined in \eqref{eq:active-sterile mixing}.
The mass-basis interactions make the connection between the active-sterile mixing parameter and observable production and decay rates explicit.
For the weak \CC,
\begin{align}
  \mathcal L_{\CC} & = \frac{g}{\sqrt2} W_\mu^- j_{W^+}^\mu + \hc ,
                   &
  j_{W^+}^\mu      & \supset \widebar\ell_\alpha \gamma^\mu \nu_\alpha^\text{int} ,
\end{align}
where
\begin{equation}
  \nu_\alpha^\text{int}
  = \left(\delta_{\alpha\beta} - \frac{\theta_\alpha \theta_\beta}{s + s^2}\right) \nu_\beta
  + \theta_\alpha \frac{i N_4 + N_5}{\sqrt2 s} .
\end{equation}
The first term contains the non-unitary light-neutrino mixing matrix and the second term describes the coupling of the charged leptons to the heavy mass eigenstates.
The corresponding weak \NC interactions can be separated according to the number of heavy neutrinos involved.
The coupling among light neutrinos is
\begin{align}
  \mathcal L_{\NC} & = \frac{g}{2\cos\theta_W} Z_\mu j_Z^\mu , &
  j_Z^\mu          & \supset
  \left(\delta_{\alpha\beta} - \frac{\theta_\alpha \theta_\beta}{s^2} \right) \widebar\nu_\alpha \gamma^\mu \nu_\beta .
\end{align}
The terms that mix one light and one heavy neutrino read
\begin{equation}
  j_Z^\mu \supset \frac{\theta_\alpha}{\sqrt2 s^2} \widebar\nu_\alpha \gamma^\mu (iN_4 + N_5) + \hc .
\end{equation}
Finally, the \NC interaction involving two \HNLs is
\begin{equation}
  j_Z^\mu \supset
  \frac{\abs{\vec \theta}^2}{2 s^2} (-i\widebar N_4 + \widebar N_5) \gamma^\mu (iN_4 + N_5) .
\end{equation}
The Higgs interactions are
\begin{align}
  \mathcal L_{NH} & = \frac{m_N}{2vs^2} h j_h ,                                                                                                                               &
  j_h             & = \theta_\alpha \widebar{\nu_\alpha^c} (i N_4 - N_5) - \frac{\abs{\vec \theta}^2}{\sqrt 2} \left(\widebar{N_4^c} N_4 + \widebar{N_5^c} N_5\right) + \hc .
\end{align}

\subsection{Mass spectrum in the presence of \LNlong violation} \label{sec:mass spectrum}

Under the assumption that the Lagrangian \eqref{eq:LSPSS} describes the physics of a single pseudo-Dirac pair, the mass splittings generated by the small amount of \LN violation present in the \LN-breaking part of the mass matrix \eqref{eq:mass matrix} depend only on the parameters listed in \eqref{eq:mass hierarchy}.

The degeneracy between the two heavy Majorana \DOFs is broken, turning the pair into a pseudo-Dirac particle.
The masses of the two heavy states $n_{\nicefrac45}$ read, to \LO in the \LN-violating mass parameters,
\begin{align} \label{eq:heavy mass splitting}
  m_{\nicefrac45}^{} & = m_N^{} \mp \frac12 \Delta m ,                                                                             &
  \Delta m           & = \abs*{\frac{2 \vec \mu_D^\dagger \vec \theta + \mu_M^*}{s^2} + \mu_M^\prime} + \order*{\frac{\mu^2}{m}} .
\end{align}
While one light neutrino remains exactly massless, the other two generally acquire masses proportional to the \LN-violating parameters.
These masses are determined by the measured mass-squared differences $\Delta m_{ij}^2 \equiv \abs{m_i^2 - m_j^2}$.
For \NO and \IO they are given by \cite{Esteban:2024eli, NuFIT:2025web}
\begin{subequations} \label{eq:light neutrino masses}
  \begin{align}
    \NO   & :                                     &
    m_1^2 & = 0 ,                                 &
    m_2^2 & = \Delta m_{21}^2 ,                   &
    m_3^2 & = \Delta m_{3k}^2 ,                   &
    k     & = 1 ,
    \\
    \IO   & :                                     &
    m_3^2 & = 0 ,                                 &
    m_1^2 & = \Delta m_{3k}^2 - \Delta m_{21}^2 , &
    m_2^2 & = \Delta m_{3k}^2 ,                   &
    k     & = 2 ,
  \end{align}
\end{subequations}
where $\Delta m_{21}^2$ and $\Delta m_{3k}^2$ denote the solar and atmospheric mass-squared splittings, respectively; they correspond to the smallest and largest of the three independent neutrino mass-squared differences.

\begin{table}
  \begin{tabular}{
      r
      c S[table-format=4.0] S[table-format=2.1,table-auto-round] S[table-format=2.3,round-mode=figures,round-precision=3]
    }\toprule
          & {$\Delta m_{21}^2 / \unit{meV^2}$} & {$\Delta m_{3k}^2 / \unit{meV^2}$} & {$\ev{m_{2j}} / \unit{meV}$} & {$\Delta m_{2j} / \unit{meV}$} \\\midrule
    $\NO$ & \multirow{2}{*}{\qty{75.37}}       & 2511                               & 29.3957                      & 41.4283                        \\
    $\IO$ &                                    & 2483                               & 49.4487                      & 0.762104                       \\
    \bottomrule
  \end{tabular}
  \caption[Light neutrino masses in the \SPSSlong]{
    Experimentally measured light-neutrino mass-squared differences \cite{Esteban:2024eli, NuFIT:2025web}, together with the corresponding mean mass and mass splitting of the two massive light neutrinos in the \SPSS \eqref{eq:neutrino-masses}, for \NO and \IO.
  }
  \label{tab:neutrino-masses}
\end{table}

The average and difference of the two non-vanishing light neutrino masses in the \SPSS are, to \LO in the \LN-violating parameters,
\begin{align} \label{eq:neutrino-masses}
  \ev{m_{2j}}   & \equiv \frac{m_2 + m_j}{2} = \abs{\vec \theta} \abs{\vec \mu_\nu}
  + \order*{\frac{\mu^2}{m}}
  ,             &
  \Delta m_{2j} & \equiv \abs{m_2 - m_j} = 2 \abs{\vec \theta^\dagger \vec \mu_\nu}
  + \order*{\frac{\mu^2}{m}}
  ,
\end{align}
where
\begin{align}
  \vec \mu_\nu & = \frac{\mu_\nu \vec \theta - \vec \mu_D}{s}
  ,            &
  \mu_\nu      & = \frac{\mu_M}{2 s} + \frac{\vec \theta^\dagger \vec \mu_D}{s+s^2}
  .
\end{align}
The experimentally observed mass-squared differences appearing in \eqref{eq:light neutrino masses} are therefore
\begin{subequations}
  \begin{align}
    \NO                    & :                                       &
    \sqrt{\Delta m_{21}^2} & = \ev{m_{2j}} - \frac12 \Delta m_{2j} , &
    \sqrt{\Delta m_{3k}^2} & = \ev{m_{2j}} + \frac12 \Delta m_{2j} , &
    j                      & = 3 ,
    \\
    \IO                    & :                                       &
    \Delta m_{21}^2        & = 2 \ev{m_{2j}} \Delta m_{2j} ,         &
    \sqrt{\Delta m_{3k}^2} & = \ev{m_{2j}} + \frac12 \Delta m_{2j} , &
    j                      & = 1 .
  \end{align}
\end{subequations}
The numerical values of the quantities appearing in the above equation are summarised in \cref{tab:neutrino-masses}.

The impact of \LN violation on the mass spectrum is captured in \cref{fig:mass splitting}.
Since the three sources of \LN violation in \eqref{eq:mass matrix} are independent and need not have the same order of magnitude, it is instructive to consider them individually.
\begin{itemize}
  \item
        The parameter $\mu_M^\prime$, while contributing to the mass splitting of the heavy neutrinos \eqref{eq:heavy mass splitting}, cannot generate tree-level light neutrino masses.

  \item
        If \LN violation is restricted to the other Majorana mass parameter,
        \begin{align} \label{eq:ISS}
          \mu_M^{} & \neq 0 , & \abs{\vec \mu_D^{}} & = \mu_M^\prime = 0 ,
        \end{align}
        the Lagrangian \eqref{eq:LSPSS} captures the low-energy effects of the inverse seesaw.
        In particular, it only generates a single light neutrino mass, such that the light and heavy neutrino mass splittings are related by
        \begin{align} \label{eq:inverse seesaw mass splittings}
          \eval*{\frac{m_\nu}{\Delta m}}_{\mu_M^\prime = \abs{\vec \mu_D^{}} = 0} & = \abs{\vec \theta}^2 ,         &
          m_\nu                                                                   & \equiv \sqrt{\Delta m_{3k}^2} .
        \end{align}

  \item
        If, instead, \LN violation is restricted to the second Yukawa interaction,
        \begin{align}
          \abs{\vec \mu_D^{}} & \neq 0 , & \mu_M^{} & = \mu_M^\prime = 0 ,
        \end{align}
        the Lagrangian \eqref{eq:LSPSS} describes the low-energy effects of the minimal linear seesaw.
        In this case the splitting between the two massive light neutrinos is related to the \HNL mass splitting by
        \begin{equation}
          \eval*{\frac{\Delta m_{2j}}{\Delta m}}_{\mu_M^\prime = \mu_M^{} = 0} = 1.
        \end{equation}
        Furthermore, the mass splittings in \eqref{eq:neutrino-masses} are suppressed when $\vec{\theta}$ and $\vec{\mu}_D$ are approximately orthogonal, a configuration naturally associated with the \IO.
        Conversely, they are enhanced when $\vec{\theta}$ and $\vec{\mu}_D$ are approximately parallel, as expected in the \NO.
\end{itemize}

\subsection{\sentence\NNOslong} \label{sec:oscillations}

\resetacronym{NNO}

\begin{figure}
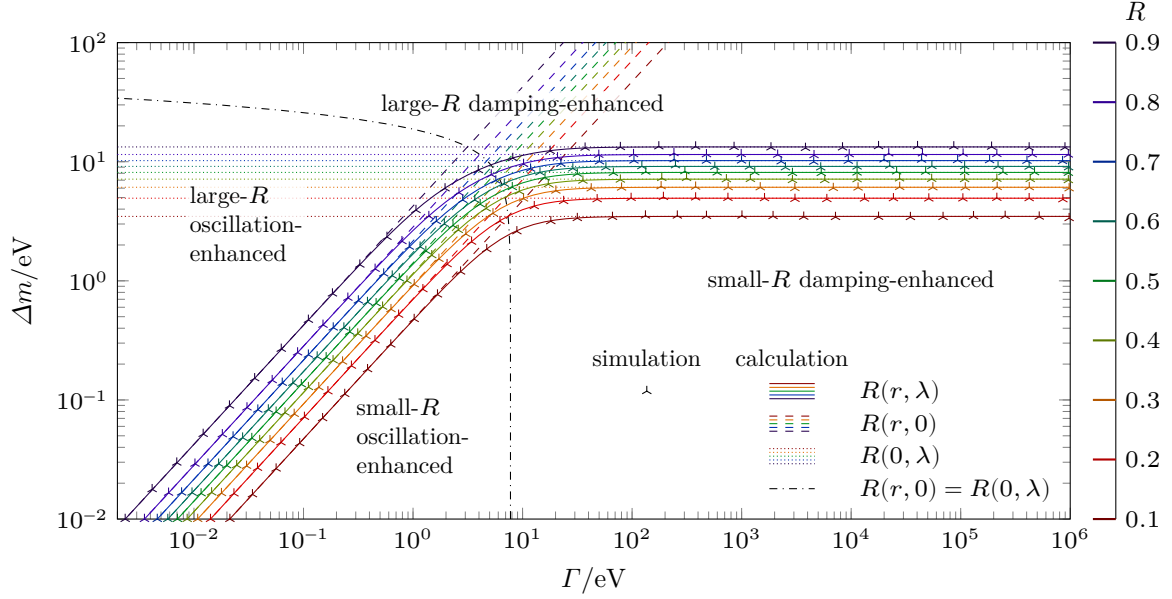

  \includepgf{Rll}
  \caption[\sentence\LNlong violation as a function of the $\HNL$ decay width and mass splitting]{
    Impact of decoherence on the \LN ratio in the range from 0.1 to 0.9.
    The simulation results of \cite{Antusch:2023nqd} (stars) are compared with the analytic expression for undamped oscillations (dashed lines), oscillations damped proportionally to the square of the mass splitting using \eqref{eq:damping threshold} with the best-fit damping threshold of \qty{7.757}{eV} (solid lines), and damping-dominated oscillations using \eqref{eq:R regimes} (dotted lines).
    The dash-dotted black line indicates the regime boundary \eqref{eq:regime boundary}.
    The four enhancement regimes introduced in \cref{sec:regimes} are separated by the \LN-ratio band and the regime boundary.
  } \label{fig:damping regimes}
\end{figure}

The small mass splitting of the pseudo-Dirac pair \eqref{eq:heavy mass splitting} leads to \NNOs that induce an oscillatory pattern of \LN-conserving and -violating events.
Without taking decoherence effects into account, the probability to measure \LN-conserving and -violating events at proper time $\tau$ is
\begin{equation} \label{eq:oscillations}
  P_\text{osc}^\pm(\tau) = \frac{1\pm \cos(\Delta m \tau)}{2} .
\end{equation}
The independent probability density function describing the decay of an unstable particle is
\begin{equation}
  f_\text{dec}(\tau) = \Gamma_N e^{-\Gamma_N\tau} ,
\end{equation}
such that the probability for decaying oscillations reads
\begin{equation}
  P^\pm(\tau) = f_\text{dec}(\tau) P_\text{osc}^\pm(\tau) .
\end{equation}
The integral over the proper time results in the total probability to observe \LN-conserving and -violating events
\footnote{
  In real experiments, the integration limits are set by the detector geometry \cite{Antusch:2022ceb}.
}
\begin{equation}
  P^\pm = \int_0^\infty P^\pm(\tau) \d \tau .
\end{equation}
The ratio of these two integrated probabilities
\begin{align} \label{eq:LN ratio theory}
  R & = \frac{P^-}{P^+} = \frac{r}{r+2}
  = \begin{cases}
      0 + \frac r2 + \order*{r^2}    & \text{for } \Delta m \ll \Gamma_N , \\
      1 - \frac 2r + \order*{r^{-2}} & \text{for } \Delta m \gg \Gamma_N , \\
    \end{cases}
    &
  r & = \left(\frac{\Delta m}{\Gamma_N}\right)^2 ,
\end{align}
is a sigmoid function, as shown in \cref{fig:damping regimes}.
This \LN ratio can be measured from the numbers of observed events with \SS and \OS charged leptons
\begin{equation} \label{eq:LN ratio}
  R \equiv \frac{N_{\SS}}{N_{\OS}} .
\end{equation}
This ratio interpolates between the two symmetry limits: an unbroken \LN-like symmetry corresponds to a vanishing \LN ratio, whereas a ratio of one represents maximal observable \LN violation.
Realistic scenarios with partial \LN violation are therefore characterised by intermediate \LN ratios.
In the symmetry-preserving limit, the pseudo-Dirac \HNL is equivalent to an exact Dirac \HNL.
In contrast, for maximal \LN violation, it decomposes into two independent Majorana \HNLs.
We refer to this regime as the double-Majorana limit.

While the cross sections describing the rates of \OS and \SS dilepton events depend on the \LN ratio, their sum is independent of it,
\begin{equation}
  \sigma \equiv \sigma_{\OS}(R) + \sigma_{\SS}(R) .
\end{equation}
Therefore, the partial cross sections in the Dirac and maximally observable \LN-violating limits satisfy
\begin{align}
  \sigma_{\OS}(0) & = \sigma ,                             &
  \sigma_{\SS}(0) & = 0 ,                                  &
  \sigma_{\OS}(1) & = \sigma_{\SS}(1) = \frac{\sigma}{2} .
\end{align}
For general values of the \LN ratio, the cross sections can be normalised to their values in the maximally \LN-violating limit
\begin{align} \label{eq:zeta}
  \zeta(R)                   & \equiv \frac{\sigma(R)}{\sigma(1)} , &
  \zeta_{\nicefrac\OS\SS}(R) & = 1 \pm \frac{1 - R}{1 + R} .
\end{align}
Standard \MC simulations of collider \HNL searches do not include \NNOs and their damping as dynamical propagation effects.
Therefore, experimental interpretations usually rely on one of the two limiting descriptions: a Dirac \HNL, for which \LN-violating rates vanish, or a Majorana \HNL, for which \LN-conserving and -violating rates are equal.

For \LN-blind searches, which combine \LN-conserving and -violating contributions, the inclusive signal rate is independent of the \LN ratio.
The dependence on the pseudo-Dirac nature of the \HNL can then enter only through effects that change acceptances or event shapes, for instance through kinematic distributions or oscillation-sensitive observables \cite{Antusch:2023jsa, Antusch:2024otj, Antusch:2026xxx}.

Searches that explicitly target \LN violation are more directly affected.
A Majorana interpretation corresponds to the double-Majorana limit and can therefore overestimate the signal rate whenever oscillations or incomplete decoherence suppress \LN violation, while a Dirac interpretation corresponds to a vanishing \LN ratio and misses any nonzero \LN-violating contribution.
Intermediate pseudo-Dirac regimes with a \LN ratio between zero and one must thus be treated by rescaling the \OS and \SS components with \eqref{eq:zeta}.

\subsection{Damping due to decoherence} \label{sec:damping}

\begin{figure}
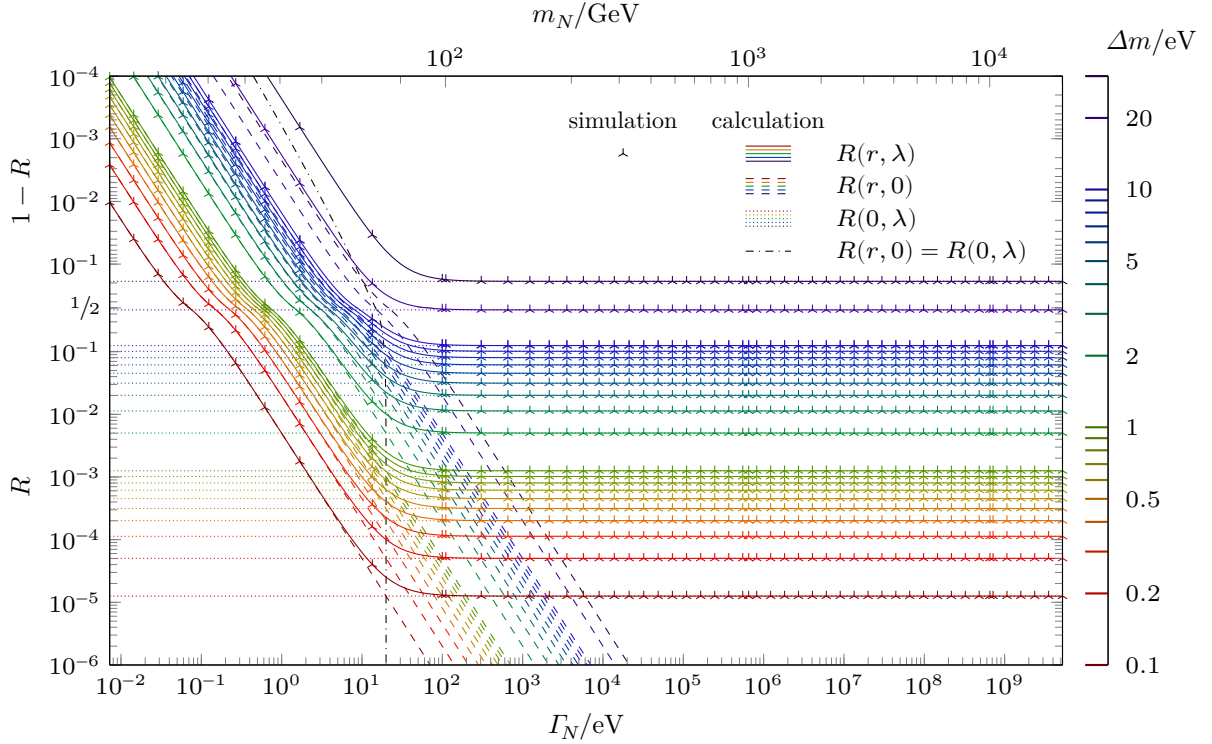

  \includepgf{Rll-mass}
  \caption[\sentence\LNlong violation as a function of the $\HNL$ mass spectrum]{
    The \LN ratio \eqref{eq:damped R} as a function of the \HNL decay width, the \HNL mass, and the mass splitting for a fixed squared active-sterile mixing of $10^{-6}$ and a damping threshold of \qty{19.94}{eV}.
    The limiting cases of pure oscillations and damping \eqref{eq:R regimes} are indicated by dashed and dotted lines, respectively.
    The regime boundary \eqref{eq:regime boundary} at which they intersect is indicated by a dash-dotted line.
  } \label{fig:Rll as function of mass}
\end{figure}

Decoherence can damp quantum oscillations.
This effect can be calculated in the framework of \QFT with external wave packets \cite{Beuthe:2001rc}.
Although decoherence arises on an event-by-event basis, its overall effect can be captured by an exponential damping of the \NNOs \eqref{eq:oscillations}
\begin{equation} \label{eq:damped oscillations}
  P_\text{osc}^\pm(\tau) = \frac{1\pm e^{-\lambda} \cos(\Delta m \tau)}{2} ,
\end{equation}
where $\lambda$ is the effective damping parameter \cite{Antusch:2023nqd}.
In the presence of damping, the \LN ratio \eqref{eq:LN ratio} becomes
\begin{equation} \label{eq:damped R}
  R = 1 - \frac{2}{1 + (1 + r) \exp\lambda}
  \to
  \begin{cases}
    \frac{r}{r + 2} & \text{for } \lambda \to 0 ,      \\
    1               & \text{for } \lambda \to \infty .
  \end{cases}
\end{equation}
Therefore, decoherence can strongly enhance the amount of observable \LN violation.
When the decay width and mass splitting are widely separated, the ratio becomes independent of the decay width at \LO,
\begin{equation} \label{eq:R limit}
  R = \begin{cases}
    \tanh \frac\lambda2 + \frac{1}{1 + \cosh \lambda} r + \order*{r^2} & \text{for } \Delta m \ll \Gamma_N , \\
    1 - 2 \frac{e^{-\lambda}}{r} + \order*{r^{-2}}                     & \text{for } \Delta m \gg \Gamma_N .
  \end{cases}
\end{equation}
The hyperbolic tangent shows that, also in this limit, the transition from minimal to maximal \LN violation follows a sigmoid function; see also \cref{fig:damping regimes}.
These limiting behaviours are demonstrated in \cref{fig:Rll as function of mass}.

The corresponding normalised cross sections \eqref{eq:zeta} read
\begin{equation} \label{eq:zeta limit}
  \zeta_{\nicefrac\OS\SS}(R) = 1 \pm \frac{1}{1 + r} e^{-\lambda} =
  \begin{cases}
    1 \pm e^{-\lambda} \mp r e^{-\lambda} + \order*{r^2} & \text{for } \Delta m \ll \Gamma_N , \\
    1 \pm \frac{e^{-\lambda}}{r} + \order*{r^{-2}}       & \text{for } \Delta m \gg \Gamma_N .
  \end{cases}
\end{equation}

The effective damping parameter has been calculated from first principles \cite{Antusch:2023nqd}.
At the \LHC, the result can be modelled as
\begin{align} \label{eq:damping threshold}
  \lambda          & = \left(\frac{\Delta m}{\Delta m_\lambda}\right)^2 ,   &
  \Delta m_\lambda & = \frac{\sigma_0}{\sigma_p} \qty{7.757 +- 0.017}{eV} , &
  \sigma_0         & = \qty{100}{nm} ,
\end{align}
where $\Delta m_\lambda$ defines the damping threshold at which decoherence alone yields an order-one contribution to the \LN ratio, and $\sigma_p$ is the wave packet width of the incoming protons.

The fit of the ratio \eqref{eq:damped R} to the \MC simulation results is shown in \cref{fig:damping regimes}.
The decay widths in the parameter space studied here range from several \unit{meV} to several \unit{TeV}.
We assume that the behaviour shown in \cref{fig:damping regimes} extends over this full range \cite{Antusch:2023nqd}.

The uncertainty in the damping threshold \eqref{eq:damping threshold} contains only the statistical error of the fit and neglects additional systematic uncertainties.
In particular, this analysis depends on assumptions regarding the widths of the incoming and outgoing particles.
The effective damping parameter \eqref{eq:damping threshold} is most sensitive to variations in the wave-packet width of the incoming protons \cite{Antusch:2023nqd}, which we estimate here from the distance between protons within the beam at the \IP,
\begin{equation}
  \sigma_p = 2 r .
\end{equation}

\subsubsection{Proton distance at hadron colliders} \label{sec:proton distance}

\begin{table}
  
\begin{tabular}{rl*5cl} \toprule
                               &                         & \multicolumn{2}{c}{$\LHC$} & \multicolumn{2}{c}{$\FCChh$} & $\SppC$                             \\\cmidrule(r){3-4} \cmidrule(l){5-6}
                               &                         & nominal & HL               & initial & nominal                                                  \\\midrule
  $\COM$ energy                & $\sqrt{s}$              & \multicolumn{2}{c}{14}     & \multicolumn{2}{c}{$100$}    & $125$   & \unit{TeV}                \\
  integrated luminosity target & $\mathcal L$            & $0.3$   & $3$              & \multicolumn{2}{c}{$30$}     & $19.5$  & \unit{ab^{-1}}            \\
  \cmidrule{1-8}
  normed transverse emittance  & $\epsilon_n$            & $3.75$  & $2.5$            & \multicolumn{2}{c}{$2.2$}    & $1.2$   & \unit{\mu m}              \\
  $\IP$ $\beta$-function       & $\beta^*$               & $55$    & $15$             & $110$   & $30$               & $50$    & \unit{cm}                 \\
  $\RMS$ $\IP$ spot size       & $\sigma$                & $16.7$  & $7.1$            & $6.8$   & $3.5$              & $3$     & \unit{\mu m}              \\
  $\RMS$ bunch length          & $l$                     & \multicolumn{2}{c}{$7.55$} & \multicolumn{2}{c}{$8$}      & $6$     & \unit{cm}                 \\
  bunch population             & $N_b$                   & $1.15$  & $2.2$            & \multicolumn{2}{c}{$1$}      & $0.4$   & $10^{11}$                 \\
  \cmidrule{1-8}
  peak luminosity              & $L$                     & $1$     & $5$              & $5$     & $30$               & $4.3$   & \qty{e34}{cm^{-2} s^{-1}} \\
  $\IP$ proton distance        & $2r$                    & $103$   & $47$             & $60.5$  & $38.9$             & $43.3$  & \unit{nm}                 \\
  \cmidrule{1-8}
  damping threshold     & $\Delta m_\lambda$      & $7.531$ & $16.50$          & $12.82$ & $19.94$            & $17.91$ & \unit{eV}                 \\
  \bottomrule
\end{tabular}

  \caption[Hadron collider beam parameters]{
    The proton distance within the beam at the \LHC, the \FCChh, and the \SppC, and therefore the damping threshold, depend on the respective beam parameters.
    The input values are taken from \cite{FCC:2018vvp, CEPCStudyGroup:2023quu}, and the dependent parameters are calculated using the equations in \cref{sec:proton distance} and \eqref{eq:damping threshold}.
  } \label{tab:beam parameter}
\end{table}

To compare the wave packet width of the protons at the \LHC and future hadron colliders, we use published beam parameter values \cite{FCC:2018vvp, CEPCStudyGroup:2023quu}.
The transverse beam size is the \RMS spot size at the \IP, obtained from the geometric emittance,
\begin{align}
  \sigma                             & = \sqrt{\epsilon_\text{geo} \beta^*} ,         &
  \epsilon_\text{geo}                & = \frac{\epsilon_n}{\beta_\text{rel} \gamma} , &
  \gamma = \frac{E_\text{beam}}{m_p} & = \frac{\sqrt s}{2m_p} ,
\end{align}
where $\epsilon_n$ is the normalised transverse emittance, $\beta^*$ is the \IP $\beta$-function, $\beta_\text{rel}$ is the relativistic velocity and almost one for the ultra-relativistic beams considered here, and $\gamma$ is the proton-beam Lorentz factor.
The resulting \IP spot sizes are of order $\sigma \approx \qty{10}{\mu m}$.
Together with an \RMS bunch length of $l \approx \qty{10}{cm}$, we estimate the size of the proton wave packet within the bunch by assuming that each proton occupies roughly a sphere of radius $r$, so that
\begin{equation}
  N_b V_\text{sphere}(r) \approx V_\text{cylinder}(l, \sigma) .
\end{equation}
The resulting diameters are of order $2 r \approx \qty{50}{nm}$, corresponding to a damping threshold of $\Delta m_\lambda \approx \qty{15}{eV}$.
The precise values for the \LHC, the \FCChh, and the \SppC are given in \cref{tab:beam parameter}.

\subsection{\sentence\LNlong violation regimes} \label{sec:regimes}

\begin{figure}
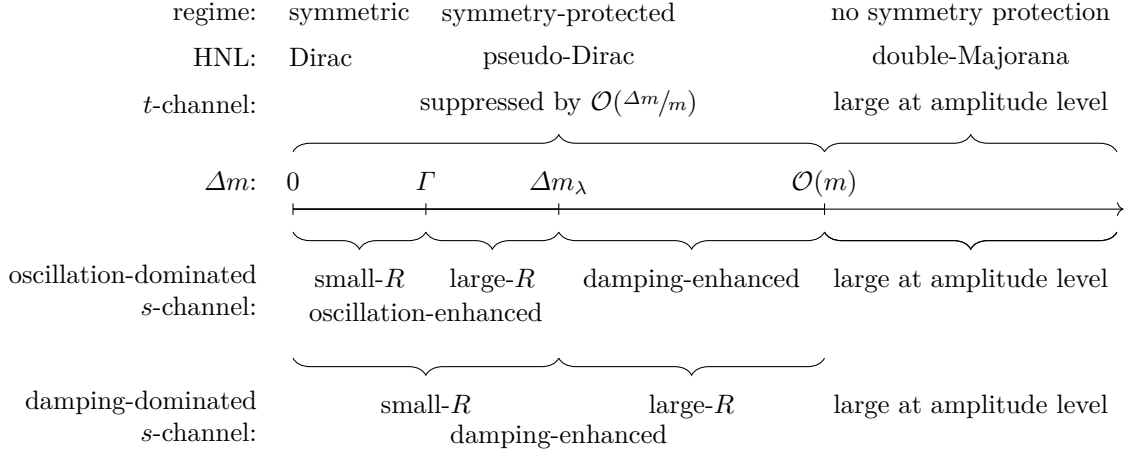

  \includepgf{lnv-energy-scale}
  \caption[\sentence\LNlong violation regimes]{
    \LN violation regimes.
    In the symmetric limit, the \HNL is a Dirac particle and \LN-violating processes are absent.
    In the symmetry-protected regime, cancellations suppress $t$-channel processes by the ratio of the mass splitting to the heavy-neutrino mass, while $s$-channel \LN violation is governed by the ratio of the mass splitting to the decay width; decoherence dominates the observable rate once the decay width exceeds the regime boundary \eqref{eq:regime boundary}.
    For very large mass splittings, the two Majorana \DOFs behave as independent particles, and both $s$-channel and $t$-channel processes exhibit unsuppressed \LN violation already at the amplitude level in plane-wave \QFT.
  } \label{fig:lnv-energy-scale}
\end{figure}

Symmetry protection, together with the emergence of \NNOs and their damping due to decoherence, gives rise to several qualitatively distinct regimes of \LN-violating physics.
The \LN-like symmetry introduced in \cref{sec:Dirac limit} enforces Dirac behaviour for the \HNL and therefore forbids any \LN-violating processes.
A slight breaking of this symmetry, as discussed in \cref{sec:mass spectrum}, permits \LN violation, but standard amplitude-level calculations in plane-wave \QFT show that it is suppressed by the ratio of the mass splitting to the heavy-neutrino mass.

While this suppression persists for $t$-channel processes, \NNOs and their damping can enhance \LN violation in $s$-channel processes, \cf \cref{sec:oscillations,sec:damping}.
To separate the dominant effects, we compare the complete \LN ratio \eqref{eq:damped R}
\begin{align}
  R(r, \lambda) & = 1 - \frac{2}{1 + (1 + r) \exp\lambda} ,            &
  r             & = \left(\frac{\Delta m}{\Gamma_N}\right)^2 ,         &
  \lambda       & = \left(\frac{\Delta m}{\Delta m_\lambda}\right)^2 ,
\end{align}
with its oscillation-dominated \eqref{eq:LN ratio theory} and damping-dominated \eqref{eq:R limit} limits,
\begin{align} \label{eq:R regimes}
  R(r, 0)       & = \frac{r}{2+r} ,       &
  R(0, \lambda) & = \tanh \frac\lambda2 .
\end{align}
Equating the two limiting contributions defines the regime boundary
\begin{align} \label{eq:regime boundary}
  \Gamma_\lambda
   & \equiv \eval*{\Gamma_N\strut}_{R(0, \lambda) = R(r, 0)}
  = \frac{\Delta m}{\sqrt{e^\lambda - 1}}
  = \Delta m_\lambda \left[1 - \frac{\lambda}{4} + \order*{\lambda^2} \right] ,
\end{align}
at which the \LN-ratio is
\begin{equation}
  \eval*{R(r, \lambda)}_{R(0, \lambda) = R(r, 0)} = \tanh\lambda .
\end{equation}

For fixed mass splitting, oscillations dominate for decay widths below this boundary, whereas damping dominates for decay widths above it.
Together with the scales set by the mass splitting and the damping threshold, this boundary separates four useful regimes, as illustrated in \cref{fig:damping regimes}:
\begin{description}
  \item[Small-\LN-ratio oscillation-enhanced regime]
        For decay widths between the mass splitting and the regime boundary, the oscillation contribution dominates while the \LN ratio remains small.
  \item[Large-\LN-ratio oscillation-enhanced regime]
        For decay widths below both the mass splitting and the regime boundary, oscillations dominate and the \LN ratio becomes large.
        The system approaches the double-Majorana limit as the mass splitting increases.
  \item[Small-\LN-ratio damping-enhanced regime]
        For decay widths above the regime boundary, damping dominates.
        If the mass splitting lies below the damping threshold \eqref{eq:damping threshold}, the \LN ratio remains small.
  \item[Large-\LN-ratio damping-enhanced regime]
        For decay widths above the regime boundary, damping dominates.
        If the mass splitting lies above the damping threshold \eqref{eq:damping threshold}, damping drives the \LN ratio to order one.
        The system approaches the double-Majorana limit as the mass splitting increases.
\end{description}
Importantly, these effects are not accessible in $t$-channel processes, which probe only the \LN violation present at the amplitude level in plane-wave \QFT.

In the regime of very large \LN violation, corresponding to mass splittings of the same order as the heavy neutrino mass, the observable \LN violation approaches that present at the amplitude level in plane-wave \QFT.
In this limit, $s$- and $t$-channel processes stand on equal footing.
However, in the absence of fine-tuning and since symmetry protection is absent within this regime, light neutrino masses are generically too large if the couplings deviate substantially from the canonical seesaw expectation.
This discussion is summarised in \cref{fig:lnv-energy-scale}.

\section{Processes} \label{sec:processes}

\dummyacronym{CC}
\dummyacronym{NC}
\dummyacronym{DY}
\dummyacronym{GF}

\begin{figure}
  \begin{panels}{.25}
    \includepgf{production-nc-dy-z}
    \caption{\DYlong} \label{fig:production:nc:DY}
    \panel{.75}
    \includepgf{production-nc-gq-1-z}\hfil
    \includepgf{production-nc-gq-2-z}\hfil
    \includepgf{production-nc-gq-3-z}
    \caption{\DYlong with initial-state radiation} \label{fig:production:nc:DYj}
    \panel{.25}
    \includepgf{production-nc-gg-z}
    \caption{\sentence \GFlong} \label{fig:production:nc:GF}
    \panel{.75}\hfil
    \includepgf{production-nc-aq-1-z}\hfil
    \includepgf{production-nc-aq-2-z}\hfil{}
    \caption{Photon-induced} \label{fig:production:nc:photon-induced}
  \end{panels}
  \caption[\sentence\NClong production channels]{
    Feynman diagrams illustrating the \NC production channels with final states $N\nu(j)$ at a proton collider.
    The \DY process is shown in panel \subref{fig:production:nc:DY}, the \DY process with an additional jet in panel \subref{fig:production:nc:DYj}, the \GF process in panel \subref{fig:production:nc:GF}, and the photon-induced process in panel \subref{fig:production:nc:photon-induced}.
    These processes do not allow for the reconstruction of \LN, since the final-state neutrino is unobservable.
  } \label{fig:production:nc}
\end{figure}

\begin{figure}
  \begin{panels}{.25}
    \includepgf{production-cc-dy}
    \caption{\DYlong} \label{fig:production:cc:DY}
    \panel{.75}
    \includepgf{production-cc-gq-1}\hfil
    \includepgf{production-cc-gq-2}\hfil
    \includepgf{production-cc-gq-3}
    \caption{\DYlong with initial-state radiation} \label{fig:production:cc:DYj}
    \panel{1}\hspace{.5em}
    \includepgf{production-cc-aq-1}\hfil
    \includepgf{production-cc-aq-2}\hfil
    \includepgf{production-cc-aq-3}\hfil
    \includepgf{production-cc-aq-4}
    \caption{Photon-induced} \label{fig:production:cc:photon-induced}
  \end{panels}
  \caption[\sentence\CClong production channels]{
    Feynman diagrams illustrating the \CC production channels with final states $N\ell(j)$ at a proton collider.
    The \DY process is shown in panel \subref{fig:production:cc:DY}, the \DY process with an additional jet in panel \subref{fig:production:cc:DYj}, and the photon-induced process in panel \subref{fig:production:cc:photon-induced}.
    Since the final state of these processes contains a charged lepton, they allow \LN-violating signatures to be measured.
  } \label{fig:production:cc}
\end{figure}

\begin{figure}
  \begin{panels}{4}
    \includepgf{decay-cc-semileptonic}
    \caption{Semi-leptonic \CC} \label{fig:decay:semileptonic:CC}
    \panel
    \includepgf{decay-cc-leptonic}
    \caption{Leptonic \CC} \label{fig:decay:leptonic:CC}
    \panel
    \includepgf{decay-nc-leptonic}
    \caption{Leptonic \NC} \label{fig:decay:leptonic:NC}
    \panel
    \includepgf{decay-nc-semileptonic}
    \caption{Semi-leptonic \NC} \label{fig:decay:semileptonic:NC}
  \end{panels}
  \caption[\sentence$\HNL$ decay modes]{
    Feynman diagrams illustrating the decay modes of \HNLs.
    The semi-leptonic \CC decay is given in panel \subref{fig:decay:semileptonic:CC},
    the leptonic \CC decay is given in panel \subref{fig:decay:leptonic:CC},
    the leptonic \NC decay is given in panel \subref{fig:decay:leptonic:NC}, and
    the semi-leptonic \NC decay is given in panel \subref{fig:decay:semileptonic:NC}.
    The decay shown in panel \subref{fig:decay:semileptonic:CC} is the only one that allows an unambiguous measurement of \LN, since the other channels produce light neutrinos.
    The decay shown in panel \subref{fig:decay:leptonic:CC} can be used to probe \LN violation, provided that the $W$ boson decays into a \LF to which the \HNL does not couple.
  } \label{fig:decay}
\end{figure}

\resetacronym{CC}
\resetacronym{NC}
\resetacronym{DY}
\resetacronym{GF}
\resetacronym{MC}

At hadron colliders, \HNLs can be produced via \NC and \CC processes.
In \NC processes, the \HNL is produced in association with a light neutrino through an intermediate $Z$ or Higgs boson.
The \NC processes considered in this work are shown in \cref{fig:production:nc}.
In \CC processes, the \HNL is produced in association with a charged lepton via an intermediate $W$ boson.
The corresponding \CC production modes are illustrated in \cref{fig:production:cc}.

The production mechanisms can be categorised as \DY processes, either with or without initial-state radiation, and photon-induced processes.
In the \NC channel, \GF production provides an additional contribution.
After propagation, the \HNL decays via semi-leptonic and fully leptonic \NC and \CC interactions, as depicted in \cref{fig:decay}.

\subsection{\sentence\LNlong-blind tri-muon final state} \label{sec:LN blind}

The \LN-blind analysis searches for fully leptonic muon processes.
The \HNLs are produced via the \CC processes shown in \cref{fig:production:cc} and subsequently decay leptonically, as shown in \cref{fig:decay:leptonic:CC,fig:decay:leptonic:NC}.
The final state is therefore
\begin{equation}
  \text{final state}(\text{\LN-blind}) = \mu^\pm\mu^\mp\mu^\pm\nu .
\end{equation}
The presence of the light neutrino in the final state prevents the reconstruction of the \LN.

Background processes mimicking this signature can be classified into two categories:
\begin{inlinelist}[itemjoin*={{ and }}]
  \item\label{item:bg:lnb:genuine} processes yielding three or more genuine leptons
  \item\label{item:bg:lnb:fake} processes in which non-leptonic objects are misidentified as prompt leptons
\end{inlinelist}

The first category \ref{item:bg:lnb:genuine}, which is the primary focus of this study, comprises processes that produce genuine leptons in the final state.
The dominant contributions arise from multi-boson production, including $3\ell\nu$, $4\ell$, $3W$, and $WW\ell\ell$, as well as top-quark processes such as $tq\ell\ell$, $t\widebar{t}\ell\nu$, and $t\widebar{t}\ell\ell$.
These processes either already resemble the signal topology or do so in configurations where the additional lepton falls outside the detector acceptance, fails identification or isolation requirements, or is otherwise not reconstructed.
They therefore require dedicated simulation and careful treatment in the analysis.

The remaining backgrounds \ref{item:bg:lnb:fake} consist of processes in which \enquote{prompt leptons} arise from misidentified hadrons, heavy-flavour decays, light meson decays, or photons within jets \cite{ATLAS:2015gtp, CMS:2015qur, CMS:2018jxx, CMS:2024xdq}.
Since \MC simulations do not reliably model lepton misidentification rates, these backgrounds are typically estimated using data-driven techniques based on control samples.
Such methods are not readily applicable to future hadron collider studies and therefore lie outside the scope of this work.
Moreover, this class of backgrounds is found to be significantly smaller than the dominant contributions in \LHC analyses \cite{CMS:2024xdq}, and is therefore neglected in the present study.

\subsection{\sentence\LNlong-violating dimuon final state} \label{sec:LN violation}

Since the unambiguous observation of \LN violation requires the full reconstruction of the final-state \LN, this analysis focuses on the \CC production channels shown in \cref{fig:production:cc}, with subsequent semi-leptonic $W$-mediated decays illustrated in \cref{fig:decay:semileptonic:CC}.
Depending on their boost, the quarks can be reconstructed as either one or two jets.
The resulting generator-level signal is therefore characterised by two \SS charged leptons accompanied by one or two jets
\begin{equation}
  \text{final state}(\text{\LN-violating}) = \mu^\pm\mu^\pm j(j) .
\end{equation}
This final state enables a direct \LN-violation measurement through the \SS lepton charges.

Background processes mimicking this signature can be classified into three categories:
\begin{inlinelist}
  \item\label{item:bg:lnv:genuine} processes yielding genuine \SS leptons in association with additional \LN-carrying particles that escape detection (\eg light neutrinos or soft charged leptons)
  \item\label{item:bg:lnv:fake} processes in which non-leptonic objects are misidentified as prompt leptons
  \item\label{item:bg:lnv:change-missid} processes producing oppositely charged leptons with misidentified charges
\end{inlinelist}

The first category \ref{item:bg:lnv:genuine} comprises processes that produce genuine \SS leptons in the final state.
The dominant channels mirror those in the \LN-blind search, now supplemented with additional hard jet activity.
These processes resemble the signal topology only in configurations where one or more additional leptons fall outside the detector acceptance, fail identification or isolation requirements, or are otherwise not reconstructed.
Consequently, they once again require dedicated simulation and careful treatment in the analysis.

The second category \ref{item:bg:lnv:fake} is analogous to its counterpart in the \LN-blind search.
While these contributions can constitute a significant fraction of the total background \cite{CMS:2018jxx} in this case, their estimation remains beyond the scope of this work.

The third category \ref{item:bg:lnv:change-missid}, while relevant for final states involving electrons \cite{ATLAS:2015gtp, CMS:2018jxx}, is negligible for muon-based searches.
Indeed, simulation studies have shown the muon charge misidentification rate to be inconsequential at the \LHC \cite{ATLAS:2015gtp, CMS:2015qur, CMS:2018jxx}.
This conclusion is supported by comparisons of charge measurements in the inner detector and muon spectrometer \cite{ATLAS:2015gtp}, as well as by studies using cosmic ray muons \cite{CMS:2009fdy}.
For this reason, and more generally because of the excellent muon momentum resolution, the present analysis focuses on \HNLs that couple exclusively to muons, rendering this background effectively zero.

\subsection{\sentence\NWAlong}

Since the processes considered here involve on-shell \HNLs that decay inside the detector, their cross sections are evaluated in the \NWA
(see \eg \cite{Abada:2022wvh}),
\begin{equation}
  \sigma( p p \to f_a (N \to f_b f_c f_d) ) = \frac{\Gamma_N(f_b f_c f_d)}{\Gamma_N} \sigma( p p \to f_a N ) ,
\end{equation}
where $\sigma( p p \to f_a N )$ is the cross section for producing an \HNL in association with the fermion $f_a$, $\Gamma_N(f_b f_c f_d)$ is the partial decay width of the \HNL into $f_b f_c f_d$, and $\Gamma_N$ is its total decay width.
To leading order in the active-sterile mixing parameters, the total \HNL decay width can be written as
\begin{equation}
  \Gamma_N = \abs{\theta_e}^2 \Gamma_N^e + \abs{\theta_\mu}^2 \Gamma_N^\mu + \abs{\theta_\tau}^2 \Gamma_N^\tau ,
\end{equation}
where the flavour coefficients are independent of the active-sterile mixing parameters.
For sufficiently large \HNL masses, the kinematic differences between the three charged-\LFs become negligible, so that
\begin{align}
  \Gamma_N^e & \approx \Gamma_N^\mu \approx \Gamma_N^\tau , &
  \Gamma_N   & \propto \abs{\vec \theta}^2 .
\end{align}

The resonant signal rates relevant for the searches in this study scale as
\begin{align}
  \begin{rcases}
    \sigma( p p \xrightarrow{N} \ell^\pm \ell^\pm jj )
    \\
    \sigma( p p \xrightarrow{N} \ell \ell \ell' \nu_{\ell'} )
  \end{rcases}
                                                & \propto \frac{\abs{\theta_\ell}^4}{\Gamma_N}
  = \abs{\vec \theta}^2 \abs{\epsilon_\ell}^4 , &
  \epsilon_\ell                                 & = \frac{\theta_\ell}{\abs{\vec \theta}} ,
\end{align}
where the approximation assumes the large-mass limit.
For an \HNL coupling to a single lepton generation, $\abs{\epsilon_\ell}=1$.

If instead the \HNL is exchanged in the $t$-channel, its momentum is spacelike and the \NWA does not apply.
The corresponding cross section is then quartic in the active-sterile mixing parameters, rather than being enhanced by the resonant propagator.
Such contributions are therefore strongly suppressed in the small-mixing regime.

\subsection{Bounds from searches at the $\LHC$} \label{sec:LHC}

Existing \LHC searches constrain several final states relevant to this work, but their published limits must be translated before they can be applied to \SPSSs.
The usual Dirac and (double-)\allowbreak Majorana benchmark interpretations summarised in \cite{ATLAS:2025qbs} correspond to limiting cases of the pseudo-Dirac parameter space, whereas an \SPSS generally yields channel-dependent rates between these limits.
The relevant distinction is not only the flavour assumption on the active-sterile mixing parameter, but also whether the analysis is \LN-blind and whether the \HNL appears as a resonant $s$-channel state or \LN violation is probed through a $t$-channel exchange.
This last distinction is central in symmetry-protected models, where observable \LN violation in resonant production and decay is controlled by \NNOs and decoherence, whereas $t$-channel amplitudes remain suppressed in the symmetry-protected regime, \cf \cref{sec:regimes}.

The most straightforward bounds come from trilepton analyses in which the \HNL decays leptonically and a light neutrino remains in the final state.
The \CMS search for final states with electrons, muons, and hadronically decaying $\tau$ leptons is the most relevant example \cite{CMS:2024xdq}.
Since the final-state neutrino carries away unobserved \LN, these limits constrain the total rate for \HNL production and decay rather than the amount of observable \LN violation.
They are therefore expected to also apply in the pseudo-Dirac regime where the \SS rate is suppressed, up to differences in flavour assumptions and acceptances.

Bounds from \LN-violating searches with resonant \HNL production are subject to the same limitations as the \SS dimuon search for \LN violation described in \cref{sec:LN violation} and therefore require reinterpretation.
This class includes the \SS dilepton searches by \ATLAS and \CMS \cite{ATLAS:2015gtp, CMS:2015qur, CMS:2018jxx}, as well as low-mass searches in on-shell $W$ and top decays \cite{ATLAS:2019kpx, ATLAS:2024fcs, ATLAS:2025lva}.
These analyses are usually interpreted in a (double-)\allowbreak Majorana benchmark model, corresponding to maximal observable \LN violation.
In the \SPSS, this assumption overestimates the \SS signal yield whenever the mass splitting is small enough that the \LN-violating rate is reduced by the \SS scaling factor \eqref{eq:zeta}.
The experimental limits from these searches should therefore be regarded as bounds on the maximally \LN-violating reference point.

By contrast, \SS $WW$ scattering or vector boson fusion searches with $t$-channel \HNL exchange \cite{CMS:2022hvh, ATLAS:2023tkz} do not directly constrain the \LN-violating search proposed here in the symmetry-protected regime.
Although their reconstructed objects, in particular \SS dimuons with jets, are close to the final state considered in \cref{sec:LN violation}, their \LN-violating amplitude is not enhanced by the propagation and decay of an on-shell \HNL.
Instead, the contributions of the two pseudo-Dirac mass eigenstates cancel up to symmetry-breaking effects, so the $t$-channel rate remains suppressed by the small \LN-violating parameters.
These limits become comparable to $s$-channel Majorana limits only for large mass splittings, where the symmetry protection is lost, see \cref{sec:regimes}.
They are therefore shown separately and should not be used to exclude the small-splitting region targeted by the present analysis.

The flavour assumptions entering the experimental limits require additional care.
The analysis in this work assumes \HNL couplings only to muons.
Searches involving electrons, $\tau$ leptons, or mixed-flavour categories constrain different combinations of active-sterile mixings and can also change the total \HNL decay width.
\LF violation and \LN violation are therefore not interchangeable: mixed-flavour final states can improve background rejection and probe the flavour structure of the mixings, but their \SS rate is still controlled by the \LN-violating suppression in \eqref{eq:zeta}.
Consequently, bounds derived under exclusive-flavour or benchmark-flavour assumptions should only be mapped onto the muon-only scenario after accounting for the corresponding production rates, branching ratios, and width effects.

The strongest constraints on Majorana \HNLs coupling exclusively to muons are derived from searches for the $\mu^\pm\mu^\pm e^\mp$ and $\mu^\pm\mu^\pm\tau^\mp$ final states at the \LHC \cite{ATLAS:2019kpx, CMS:2024xdq, ATLAS:2025lva}.
Under this coupling assumption, these signatures can only originate from \LN-violating processes and are therefore suppressed by the ratio \eqref{eq:zeta}.
When other flavour couplings are present, however, the same final states can arise from unsuppressed \LN-conserving but \LF-violating decays.
Consequently, although the absence of an excess constrains \LN-violating models, the observation of one would not constitute unambiguous evidence for either \LN or \LF violation.

\section{\MClong event generation} \label{sec:event generation}

Parton-level events are generated with \MG \cite{Alwall:2014hca}.
The resulting events are subsequently passed to \software[8.316]{Pythia} \cite{Bierlich:2022pfr} for parton showering and hadronisation, and then processed with the fast detector simulator \software[3.5]{Delphes} \cite{deFavereau:2013fsa}.
The latter employs a modified \FCChh detector card.\footnote{
  To prevent double counting of muons, we have implemented the adjustments discussed in \cite{Delphes:Issue:152}.
}

\PDFs are included via \software[6.5.5]{LHAPDF} \cite{Andersen:2014efa, Buckley:2014ana} through the corresponding interface in \MG.
All simulations use the NNPDF3.1luxQED $\NLO$ \PDF set (\code{LHAPDF ID = 324900}) \cite{Bertone:2017bme}, which incorporates the photon content of the proton as predicted in the LUXqed formalism \cite{Manohar:2016nzj, Manohar:2017eqh}.
During event generation, the \MG beam ID codes \code{lpp1} and \code{lpp2} are kept at their default value (\code{1 = proton}) \cite{Pascoli:2018heg}.

The parton simulations use the \MG multiparticle definitions
\begin{verbatim}
define p  = g d d~ u u~ s s~ c c~ b b~  define j  = g d d~ u u~ s s~ c c~ b b~
define qq = u u~ c c~ d d~ s s~ b b~    define ww = w+ w-
define ll = mu+ mu-                     define vv = vm vm~
define l- = mu-                         define l+ = mu+
define v  = vm                          define v~ = vm~
define tt = t t~                        define bb = b b~
\end{verbatim}
For consistency, these definitions also require setting \code{maxjetflavor} to \code{5}.
In the following, we use them to provide the syntax for a generic leptonic analysis, although our study focuses exclusively on final states involving muons.
For signal processes, we additionally use
\begin{verbatim}
define nn = n1 n2
\end{verbatim}

Generator-level cuts on partons can be applied within the \MG simulation.
These cuts regulate \IR and collinear divergences and reduce the number of generated events that would otherwise be discarded during the subsequent \software{Delphes} detector simulation and event reconstruction.
The following generator-level cuts are imposed
\begin{subequations} \label{eq:preselection cuts}
  \begin{align}
    p_T^{}(\ell)         & > \qty{18}{GeV} , &
    \abs{\eta(\ell)}     & < 6.5 ,           &
    \Delta R(\ell, \ell) & > 0.2 ,
    \\
    p_T^{}(q)            & > \qty{20}{GeV} , &
    \abs{\eta(q)}        & < 6.5 ,           &
    m(q, q)              & > \qty{20}{GeV} .
  \end{align}
\end{subequations}
To compute total production cross sections, these cuts are omitted in \cref{fig:production:xsection,tab:signal-syntax}, with the exception of the \DY process with additional jets, in which a transverse jet momentum cut of \qty{1}{GeV} is retained.

Where necessary, MLM matching \cite{Alwall:2007fs} is enabled in \MG with
\begin{verbatim}
set ickkw 1                set pdfwgt T                set xqcut 20
\end{verbatim}
to avoid double counting jets produced by \MG and \software{Pythia}.
For technical reasons, matching prevents the inclusion of the minimum separation $\Delta R(\ell, q)$ in the cuts \eqref{eq:preselection cuts}.

To perform \MC simulations, the underlying models must first be implemented in \software{FeynRules}.
The implementations are then processed with \software[2.3]{FeynRules} \cite{Christensen:2008py, Alloul:2013bka} to generate \UFO files suitable for one-loop simulations.
Subsequently, they are imported into \MG for event generation.

The background processes are simulated using the \SM \UFO files distributed with \MG, with the \code{no_masses} restriction applied to treat light fermions as massless.
The signal simulations employ a new \software{FeynRules} model developed for this purpose.
For all results, we assume that the \FCChh runs with a \COM energy of \qty{100}{TeV} and reaches a combined integrated luminosity of \qty{30}{ab^{-1}} \cite{FCC:2018vvp}.

\subsection{\sentence\SPSSlong implementation}

To model the collider phenomenology of the \SPSS, we developed a dedicated \software[1.0]{SPSS} \cite{FR:SPSS} implementation in \software{FeynRules}, superseding the previously used \software{pSPSS} model \cite{Antusch:2022ceb, Antusch:2023jsa, FR:pSPSS}.
In contrast to the earlier approach, the \software{SPSS} model does not encode the mass splitting \eqref{eq:heavy mass splitting} and effective damping parameter \eqref{eq:damped oscillations} directly in the model file.

The model contains the two Majorana neutrinos \code{Na}, which form the degenerate Dirac pair of the \LN-conserving \SPSS discussed in \cref{sec:SPSS}; their mass and width are denoted by \code{mN} and \code{WN}.
The physical amount of \LN violation, arising from \NNOs and decoherence, is introduced at the analysis stage rather than through a patch to \MG.

The interactions of \HNLs with \SM particles are governed by the active-sterile mixing parameters \eqref{eq:active-sterile mixing}, which can be taken to be real without loss of generality in the \LN-conserving case
\begin{align}
  V_{1i} & \equiv \abs{\theta_{1i}} , &
  i      & = e, \mu, \tau.
\end{align}
These parameters are specified by the user in terms of \code{V1e}, \code{V1m}, and \code{V1t}.
The rotation from mass to interaction eigenstates via the neutrino mixing matrix \eqref{eq:neutrino mixing matrix} is implemented exactly to all orders in $\theta_{1i}$.

\subsection{Signal}

\begin{figure}
  \includepgf{production-xsection}
  \caption[\sentence$\HNL$ production cross sections]{
    \HNL production cross sections for the \NC processes presented in \cref{fig:production:nc} and the \CC processes presented in \cref{fig:production:cc}.
    For light \HNLs, both production modes are dominated by \DY processes.
    In the \NC channel, the \GF contribution dominates for \HNL masses around \qty{100}{GeV} and becomes relevant once again for masses above \qty{1}{TeV}.
    For \CC processes, the photon-induced channel overtakes \DY and becomes dominant at masses around \qty{1}{TeV}.
    The simulation is performed at a squared active-sterile mixing of $10^{-6}$.
  } \label{fig:production:xsection}
\end{figure}

To generate the \NC \DY production channels depicted in \cref{fig:production:nc:DY,fig:production:nc:DYj} using the \software{SPSS} model, we employ the \MG syntax
\begin{verbatim}
generate    p p > nn vv
generate    p p > nn vv j
\end{verbatim}
The simulation of photon-induced channels depicted in \cref{fig:production:nc:photon-induced} uses \cite{Degrande:2016aje, Pascoli:2018heg}
\begin{verbatim}
generate    p a > nn vv j
add process a p > nn vv j
\end{verbatim}
To simulate the loop-induced \GF diagrams depicted in \cref{fig:production:nc:GF}, it is necessary to activate \MG's \NLO formalism \cite{Frixione:2002ik, Hirschi:2015iia, Frederix:2018nkq} through the \code{[QCD]} flag in
\begin{verbatim}
generate    g g > nn vv aS=4 [QCD]
\end{verbatim}
Analogously to their \NC counterparts, the \CC \DY and photon-induced production channels depicted in \cref{fig:production:cc:DY,fig:production:cc:DYj,fig:production:cc:photon-induced} are produced by
\begin{verbatim}
generate    p p > nn ll                  generate    p a > nn ll j
generate    p p > nn ll j                add process a p > nn ll j
\end{verbatim}

In proton-proton collisions at a \COM energy of \qty{100}{TeV}, light \HNLs are produced predominantly via \DY processes in both the \CC and \NC channels.
Including initial-state radiation in the standard \DY process further enhances the cross section.
For \HNL masses near or above the \EW boson resonances, the \DY contributions become increasingly suppressed.
In contrast, in the \NC channel, \GF peaks at masses around \qty{100}{GeV}; after becoming subdominant, it regains competitive again at higher masses.
Similarly, in the \CC channel, photon-induced processes, whose cross sections decrease more slowly than those of the \DY channels, become the dominant production mechanism for \HNL masses above \qty{1}{TeV}.
These cross sections are depicted in \cref{fig:production:xsection}.

\subsubsection{\sentence\HNLlong production and decay}

The combined simulation of \HNL production and decay required for a semi-leptonic search, \cf \cref{sec:LN violation}, can be implemented using
\begin{verbatim}
generate    p p > nn ll  , (nn > ll j j)
add process p p > nn ll j, (nn > ll j j)
add process p a > nn ll j, (nn > ll j j)
add process a p > nn ll j, (nn > ll j j)
\end{verbatim}
This setup enforces the \NWA by explicitly placing the \HNL on shell.
This step is necessary because the \software{SPSS} model is formulated in the symmetric limit, in which \LN-violating effects are absent.
Consequently, no \LN-violating events are generated by \MG.

Enforcing the on-shell condition for the \HNLs eliminates the interference between different mass eigenstates.
This reproduces the behaviour of a double-Majorana scenario, in which \LN-conserving and -violating events are produced with equal probability.
The correct ratio arising from \NNOs can subsequently be incorporated either on an event-by-event basis \cite{Antusch:2022ceb, Antusch:2024otj} or by rescaling the cross section with the \LN ratio \eqref{eq:zeta}.

\begin{table}
  \begin{tabular}{rl*{3}{S[table-format=2.1(1),exponent-mode=fixed,fixed-exponent=0,uncertainty-mode=compact,round-mode=uncertainty,round-precision=1]}} \toprule
    \multicolumn{2}{c}{syntax}                      & \multicolumn{2}{c}{\LN}                    & total               \\\cmidrule(r){1-2} \cmidrule(l){3-4}
    technique   & \header{example}                  & {conserving}        & {violating}                                \\\midrule
    on-shell    & \code{p p > nn ll, (nn > ll j j)} & 2.527(0.003109)e+01 & 2.713(0.003077)e+01  & 5.228(0.006264)e+01 \\
    $s$-channel & \code{p p > n1 | n2 > ll ll j j}  & 4.908(0.02387)e+01  & {$\approx 10^{-32}$} & 4.886(0.02407)e+01  \\
    \NPlong     & \code{p p > ll ll j j NP>1}       & 4.896(0.01586)e+01  & 0                    & 4.88(0.02215)e+01   \\
    \bottomrule\end{tabular}
  \caption[Impact of $\MG$ syntax on the cross section]{
    Comparison of the normalised production cross sections $\sigma \abs{\vec \theta}^{-2}$ given in $\unit{pb}$ for processes that are purely \LN-conserving and -violating, as well as their combination, using different \MG syntax choices.
    Each simulation is performed with an \HNL mass of \qty{100}{GeV} and a squared active-sterile mixing of $10^{-6}$.
    The on-shell syntax generates an equal amount of \LN-conserving and -violating events.
    The reported difference can be attributed to generator-level cuts that affect these two samples differently because of kinematic differences in the final-state distributions \cite{Antusch:2024otj}.
  } \label{tab:signal-syntax}
\end{table}

Alternatively, the full process can be simulated by requiring an $s$-channel \HNL or by using the \NP scale in the \software{SPSS} model, for example via
\begin{verbatim}
generate p p > n1 | n2 > ll ll j j        generate p p > ll ll j j NP>1
\end{verbatim}
The latter syntax includes diagrams with the \HNL in both the $s$- and $t$-channels; the $t$-channel contribution can be isolated using
\begin{verbatim}
generate p p > ll ll j j $$ nn NP>1
\end{verbatim}
which confirms that it is suppressed by a factor of $\abs{\vec \theta}^2$ with respect to the $s$-channel.
This makes the $t$-channel contribution negligible except for the largest mixings, even before the additional suppression from the \LN-like symmetry is taken into account.

The available syntax choices are not interchangeable and may produce different results.
These discrepancies are summarised in \cref{tab:signal-syntax}.
The syntax used in the present work does not provide a physically accurate description without subsequent corrections; it is adopted because other syntax choices produce no \LN violation within the \software{SPSS} model.

To generate specific final states, such as only \LN-violating events, \code{ll} can be selectively replaced by \code{l+} and \code{l-}.

The simulation of the processes required for a \LN-blind search, \cf \cref{sec:LN blind}, is largely analogous to the one described above, except that a charged lepton and a light neutrino replace the quarks produced in the \HNL decay.
In terms of syntax, this corresponds to replacing \code{j j} with \code{ll vv} throughout.

\subsubsection{Scaling with the active-sterile mixing}

\begin{figure}
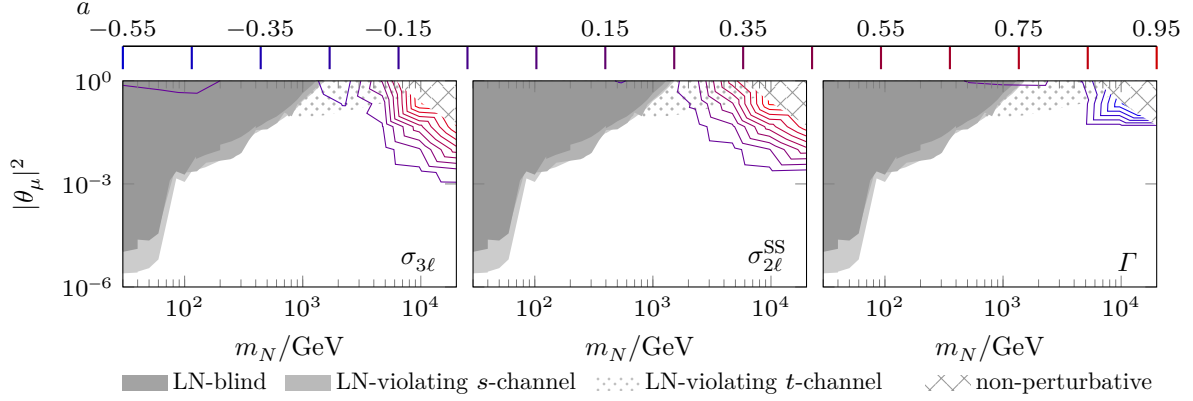

  \includepgf{scaling}
  \caption[Dependence of $\HNL$ production and decay on the active-sterile mixing]{
    Relative difference \eqref{eq:signal-scan-asymmetry} between the double-Majorana cross sections and decay widths obtained from a parameter scan over the heavy-neutrino mass and active-sterile mixing in \MG and those derived from a scan over the heavy-neutrino mass at fixed squared active-sterile mixing of $10^{-6}$, rescaled by \eqref{eq:signal-scaling}.
    The dark grey area denotes the \LHC exclusion limits from \LN-blind searches, the light grey regions represent exclusion limits derived from \LN-violating searches with \HNLs in the $s$-channel, and the dotted area indicates bounds from \LN-violating searches with \HNLs in the $t$-channel, \cf \cref{sec:LHC}.
    The hashed area indicates the non-perturbative parameter space where the \HNL Yukawa coupling satisfies $y_1 \geq 8\pi$.
  } \label{fig:signal-scaling}
\end{figure}

To ensure computational efficiency and avoid numerical instabilities at large Yukawa couplings, the \MG parameter-space scan is restricted to variations in the \HNL mass, while the squared active-sterile mixing is fixed to $10^{-6}$.
The resulting cross section as a function of the \HNL mass is shown in \cref{fig:reconstruction:xsection}.
Except for very low masses, the \LN-violating processes exhibit larger cross sections than their \LN-blind counterparts.

The remaining parameter space can be probed by rescaling the \HNL decay width and the process cross sections, collectively denoted by $x \in \{ \Gamma_N, \sigma_{3\ell}, \sigma_{2\ell}^{\SS} \}$, according to
\footnote{
  The scaling of the decay width reduces the number of displaced vertices.
  However, since such vertices are nearly absent within the parameter space under study, this effect is neglected.
}
\begin{align} \label{eq:signal-scaling}
  \frac{x(\vec \theta)}{x(\vec \theta_0)} & = \left( \frac{s_0}{s} \right)^{p_x} \frac{\abs{\vec \theta}^2}{\abs{\vec \theta_0}^2} , &
  p_{\Gamma_N}^{}                         & = 2.5 ,                                                                                  &
  p_{\sigma_{3\ell}}                      & = 3.1 ,                                                                                  &
  p_{\sigma_{2\ell}^{\SS}}                & = 1.5 ,
\end{align}
where the singular value \eqref{eq:active-sterile mixing} is defined by $s_0^2 = 1 + \abs{\vec \theta_0}^2$.
The different scaling exponents for the \LN-blind and \LN-violating channels reflect the fact that the former includes contributions from both \NC and \CC decays, whereas the latter considers only \CC decays.
Using a universal scaling factor for each search channel neglects residual channel-by-channel differences in the selection efficiency.
This is expected to have a minor impact, particularly for small active-sterile mixing angles, where the dependence on the singular values vanishes:
\begin{equation} \label{eq:signal-scaling-approximation}
  \frac{x(\vec \theta)}{x(\vec \theta_0)} \approx \frac{\abs{\vec \theta}^2}{\abs{\vec \theta_0}^2} .
\end{equation}

To quantify the difference between the full simulation and the scaling procedure, we define the relative difference as
\begin{align} \label{eq:signal-scan-asymmetry}
  a_x = \frac{x_{\text{scaled}} - x_{\MG}}{x_{\text{scaled}} + x_{\MG}} ,
\end{align}
which is bounded by $\pm1$ and vanishes when the two quantities are equal.
The agreement between the two approaches, evaluated for a subset of parameter points, is demonstrated in \cref{fig:signal-scaling}.

\subsection{Background}

For this study, we simulate only background sources that can be reliably modelled using \MC methods.
Consequently, the simulation setup described in the previous section remains largely unchanged.
The main difference is that the transverse momentum preselection cut \eqref{eq:preselection cuts} for charged leptons outside the signal signature is adapted to
\begin{equation}
  p_T^{}(\ell) < \qty{40}{GeV} ,
\end{equation}
which reduces the number of simulated events that would fail the selection criteria once the additional charged leptons are identified.

The first class of background processes arises from \EW production of three or four charged leptons,
\begin{align}
  p p & \to \ell \ell \ell \nu ,  &
  p p & \to \ell \ell \ell \ell .
\end{align}
The simulation of the $3\ell\nu$ background in \MG requires the two final states to be written explicitly \cite{MadGraph5:Question:262194}
\begin{verbatim}
generate    p p > l+ l- l+ v  / h
add process p p > l+ l- l- v~ / h
\end{verbatim}
The $4\ell$ final state can be simulated using
\begin{verbatim}
generate    p p > ll ll ll ll / h
\end{verbatim}

Despite being suppressed relative to their diboson counterparts, processes with additional gauge bosons can still be significant \cite{Pascoli:2018heg}.
In particular, the resonant production of multiple $W$ bosons that subsequently decay leptonically is relevant
\begin{align}
  p p & \to (W \to \ell \nu) (W \to \ell \nu) (W \to \ell \nu) , &
  p p & \to (W \to \ell \nu) (W \to \ell \nu) \ell \ell .
\end{align}
The \MG syntax to simulate these processes is
\begin{verbatim}
generate    p p > w+ w+ w- / h, (ww > ll vv)
add process p p > w+ w- w- / h, (ww > ll vv)
generate    p p > ww ww ll ll / h, (ww > ll vv)
\end{verbatim}

The final background category results from the production of top quarks that subsequently decay as $t \to b (W \to \ell \nu)$ alongside \EW bosons
\begin{align}
  p p & \to t q \ell \ell ,           &
  p p & \to t \widebar{t} \ell \nu ,  &
  p p & \to t \widebar{t} \ell \ell .
\end{align}
These can be simulated via the \MG syntax
\begin{verbatim}
generate    p p > tt qq ll ll / h, (tt > ww bb, ww > ll vv)
generate    p p > t  t~ ll vv / h, (tt > ww bb, ww > ll vv)
generate    p p > t  t~ ll ll / h, (tt > ww bb, ww > ll vv)
\end{verbatim}

The background processes discussed thus far are capable of mimicking the signatures of two \SS charged leptons and three charged leptons.
However, they generally fail to reproduce the hard jets expected from the semi-leptonic decay of an \HNL.
The study of \LN-violating signatures must therefore include up to two hard jets in the existing background samples.
For instance, the $4\ell(jj)$ background would be generated with
\begin{verbatim}
generate    p p > ll ll ll ll     / h
add process p p > ll ll ll ll j   / h
add process p p > ll ll ll ll j j / h
\end{verbatim}

The $tq\ell\ell$ background, which already includes one hard light quark, should be supplemented with a single additional jet.
However, an ongoing issue in the interface between \MG and \software{Pythia} limits the simulation of this process \cite{MadGraph5:Question:821634, Pythia:work_item:588}.
Accordingly, the \LN-violating search employs the $tq\ell\ell$ simulation without additional jets.

The $t\widebar{t}\ell\ell$ background should ideally include contributions with up to two additional jets.
However, because of a limitation in the simulation of this process, only configurations with a single additional jet are available.
Furthermore, to regulate collinear divergences, it is necessary to impose
\begin{equation}
  m(\ell, \ell) \geq \qty{5}{GeV} .
\end{equation}

\begin{table}
  
\begin{tabular}{
  ll
  S[table-format=1.2e+1,table-auto-round,exponent-mode=scientific]
  S[table-format=2.1,table-auto-round]
  S[table-format=1.2e1,table-auto-round,exponent-mode=scientific]
  S[table-format=1.2,round-mode=figures,round-precision=2]
  S[table-format=1.2e+1,table-auto-round,exponent-mode=scientific]
  S[table-format=2.1,table-auto-round]
  S[table-format=1.2e1,table-auto-round,exponent-mode=scientific]
  S[table-format=2.2,round-mode=figures,round-precision=2]
  }\toprule
  \multicolumn{2}{c}{process}                     & \multicolumn{4}{c}{$\LN$-blind search}                                                                                           & \multicolumn{4}{c}{$\LN$-violating search}                                                                                       \\\cmidrule(lr){3-6}\cmidrule(l){7-10}
                         &                        & $\sigma / \unit{pb}$ & $\epsilon / \unit{\%}$ & $\epsilon  \sigma  \mathcal L$ & {$\flatfrac{\sigma  \mathcal L}{n_\text{sim}}$} & $\sigma / \unit{pb}$ & $\epsilon / \unit{\%}$ & $\epsilon  \sigma  \mathcal L$ & {$\flatfrac{\sigma  \mathcal L}{n_\text{sim}}$} \\\midrule
  \multirow{4}{*}{$\EW$} & $3\ell\nu$             & 0.8232110677832601   & 68.39576000000001      & 16891243.986434277             & 4.93926640669956                                & 4.676418346495805    & 11.797165591600358     & 16550544.482762685             & 29.35742954938748                               \\
                         & $4\ell$                & 0.18866396301999994  & 42.5952                & 2410853.7712888503             & 5.6599188905999975                              & 1.3248765833190759   & 2.9748372636066818     & 1182387.6689212467             & 22.70329625424821                               \\
                         & $3W$                   & 0.0015937956         & 71.37                  & 34124.7575916                  & 0.47813868000000004                             & 0.022766365552554103 & 11.634300360697416     & 79461.22048795484              & 7.058827439633547                               \\
                         & $WW\ell\ell$           & 0.000832080967473    & 42.158                 & 10523.660828018019             & 0.24962429024189997                             & 0.07389122883433015  & 2.9947493633210014     & 66385.71315198498              & 11.61604779562292                               \\         \cmidrule(r){1-2}\cmidrule(lr){3-6}\cmidrule(l){7-10}
                         & $tq\ell\ell$           & 0.0973888064402844   & 67.08539999999999      & 1960010.1106707163             & 2.921664193208532                               & 0.12003823387131726  & 13.3193                & 479647.57452067087             & 3.601147016139518                               \\
  top                    & $t\widebar{t}\ell\nu$  & 0.0053404            & 62.985                 & 100909.52820000002             & 1.60212                                         & 0.06834324227231225  & 15.83158261399632      & 324594.50584274315             & 22.3181040870973                                \\
                         & $t\widebar{t}\ell\ell$ & 0.01465608756716     & 43.277                 & 190281.45049319498             & 4.396826270148                                  & 0.0587936079903337   & 6.250639072373668      & 110249.28699306016             & 18.035217895151344                              \\
  \bottomrule
\end{tabular}

  \caption[Background cross sections and reconstruction rates]{
    Cross section, event reconstruction rate, expected number of events, and the number of physical events per simulated event for each background source of the searches for \LN-blind and \LN-violating signals.
    The integrated luminosity of \qty{30}{ab^{-1}} assumed here corresponds to the luminosity goal of the \FCChh \cite{FCC:2018vvp}.
  } \label{tab:background-xsection}
\end{table}

The cross sections of the simulated background processes are summarised in \cref{tab:background-xsection}.
Diboson production constitutes the dominant source of background, while top-associated processes provide subleading but non-negligible contributions.
Triboson processes, while suppressed in fully leptonic channels, become comparable to top-associated backgrounds when supplemented with hard jet activity.

\section{Analysis strategy} \label{sec:analysis strategy}

\resetacronym{BDT}

The present work builds upon previous searches performed by \ATLAS \cite{ATLAS:2015gtp} and \CMS \cite{CMS:2015qur, CMS:2018jxx} at the \LHC, as well as earlier studies targeting the \FCChh \cite{Antusch:2016ejd, Antusch:2018bgr, Pascoli:2018heg}.

We develop a two-stage analysis within the \software[6.38.04]{ROOT} framework \cite{Brun:1997pa}, making use of the \software{TMVA} machine-learning library \cite{TMVA:2007ngy}.
First, a set of baseline selection criteria is applied to reconstruct the target event topology and suppress the dominant background contributions.
The events passing this preselection are then analysed using a multivariate \BDT-based approach aimed at maximising the discrimination between signal and background.

To ensure reproducibility, the complete simulation, analysis, and plotting pipeline is made publicly available in \cite{bruno_m_s_oliveira_2026_20830469}.

\subsection{Cut-based preselection}

\begin{figure}
  \begin{panels}{3}
    \includepgf{reconstruction-xsection}
    \caption{Cross section} \label{fig:reconstruction:xsection}

    \panel
    \includepgf{reconstruction-efficiency}
    \caption{Reconstruction rate} \label{fig:reconstruction:selection}

    \panel
    \includepgf{bdt-separation-semilocal}
    \caption{\BDT separation} \label{fig:reconstruction:bdt-separation}
  \end{panels}
  \caption[Comparison between the proposed searches]{
    Comparison between the searches for the \LN-blind and \LN-violating channels presented in \cref{sec:LN blind,sec:LN violation}.
    The cross section, event reconstruction efficiency, and \BDT separation \eqref{eq:BDT separation} are shown in panels \subref{fig:reconstruction:xsection}, \subref{fig:reconstruction:selection}, and \subref{fig:reconstruction:bdt-separation}, respectively.
  }\label{fig:reconstruction}
\end{figure}

In the first stage of the \LN-blind search, events containing energetic electrons are vetoed by requiring the absence of electrons with transverse momentum larger than \qty{20}{GeV}.
A clean and well-defined tri-muon final state is then selected by requiring exactly three prompt muons with a transverse momentum larger than \qty{20}{GeV}.
\footnote{
  For the purposes of this analysis, a prompt particle is defined as having an impact parameter smaller than \qty{100}{\mu m}.
  Although \LN violation from \NNOs could potentially be suppressed by this cut, the near absence of displaced vertices within the parameter space under study means this effect is negligible.
}
Although a more refined analysis, particularly vetoing certain hadronic activity, could improve the sensitivity of the \LN-blind channel \cite{Pascoli:2018heg}, such an investigation lies beyond the scope of this work.

The \LN-violating search follows a similar selection strategy, but instead requires two prompt muons with the same electric charge.
Furthermore, events are required to contain at least two reconstructed jets, thereby ensuring sufficient hadronic activity consistent with the expected signal topology.

For \HNLs with masses below \qty{200}{GeV}, the \LN-blind search retains more signal than its \LN-violating counterpart, mostly because muons near jets are removed.
In the intermediate mass range, up to approximately \qty{1}{TeV}, the \LN-violating search is slightly more efficient.
Beyond this point, the efficiencies of both searches converge and remain nearly constant at around \qty{80}{\%}.
This behaviour is demonstrated in \cref{fig:reconstruction:selection}.

The efficiencies of this initial selection in suppressing the various background sources are summarised in \cref{tab:background-xsection}.
The \LN-blind search exhibits a strong rejection of background processes featuring an excess of charged leptons.
The \LN-violating search similarly leaves mainly backgrounds with three leptons, but achieves a higher overall background suppression.
As a result, the remaining background is dominated by $3\ell\nu$ events, with smaller contributions from $4\ell$ and top processes.

\subsection{\sentence\BDTlong-based analysis} \label{sec:BDT}

\begin{table}
  \begin{panels}{2}
    \begin{tabular}{>{\codestyle}r>{\codestyle}l}\toprule
      \normalfont parameter & \normalfont value \\\midrule
      AnalysisType          & Classification    \\
      \bottomrule
    \end{tabular}
    \caption{Factory}
    \panel
    \begin{tabular}{>{\codestyle}r>{\codestyle}l}\toprule
      \normalfont parameter & \normalfont value \\\midrule
      NTrees                & 800               \\
      MaxDepth              & 3                 \\
      MinNodeSize           & 5\%               \\
      nCuts                 & 20                \\
      BoostType             & AdaBoost          \\
      AdaBoostR2Loss        & Quadratic         \\
      AdaBoostBeta          & 0.5               \\
      SeparationType        & GiniIndex         \\
      \bottomrule
    \end{tabular}
    \caption{\BDT}
  \end{panels}
  \caption[\software{TMVA} configuration options]{
    Configuration options passed to \software{TMVA}.
  } \label{tab:bdt}
\end{table}

In the second stage, the events passing the baseline selection are analysed using a multivariate \BDT approach.
The input variables are
\begin{inlinelist}
  \item the transverse momenta and pseudorapidities of the selected muons
  \item in the \LN-violating search, the transverse momentum, pseudorapidity, and invariant mass of the two hardest jets and their combination
  \item the $\Delta \phi$, $\Delta \eta$, and $\Delta R$ separation between the aforementioned reconstructed objects
  \item the total missing transverse energy
\end{inlinelist}
The configuration options passed to the \code{TMVA::Factory} constructor and the \code{BookMethod} function are summarised in \cref{tab:bdt}.

\begin{figure}
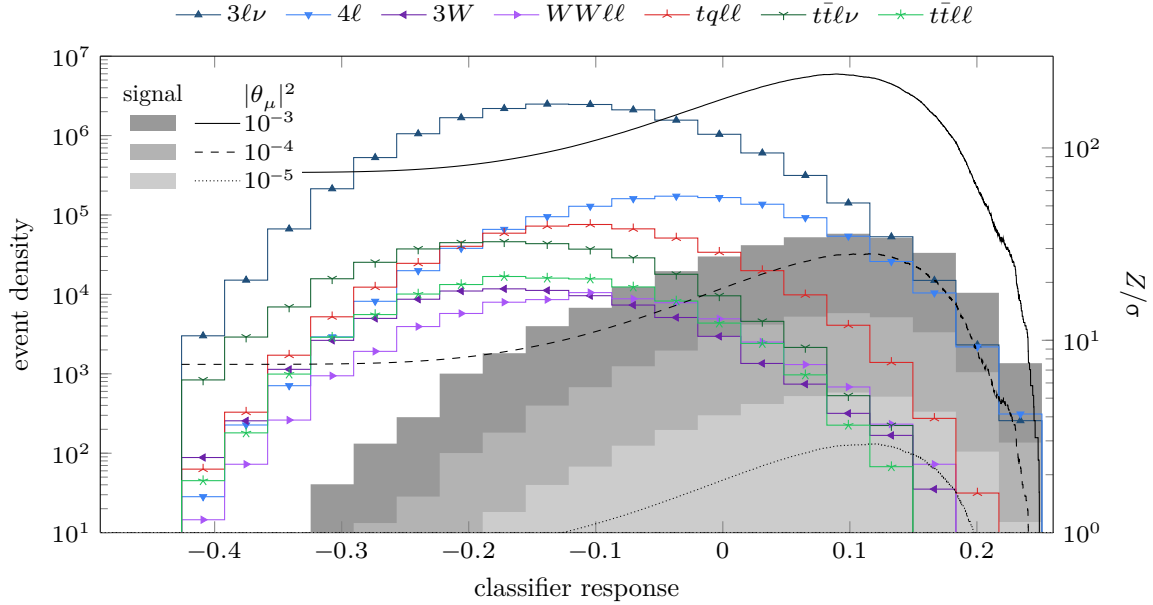

  \includepgf{bdt-distribution}
  \caption[Example \BDTlong result]{
    Distribution of signal and background evaluation events as a function of the \BDT classifier response.
    The dataset corresponds to the search for \LN violation at an \HNL mass of \qty{100}{GeV} and multiple values of the squared active-sterile mixing.
    Black curves show the exclusion significance \eqref{eq:significance-exclusion-discovery} when applying a lower threshold on the classifier response.
  } \label{fig:bdt-distribution}
\end{figure}

After training, \software{TMVA} provides, among other outputs, the distribution of events as a function of the classifier response, as depicted in \cref{fig:bdt-distribution}.
The separation between the signal and background hypotheses can be quantified in terms of two complementary measures constructed from their probability density functions $f_s(x)$ and $f_b(x)$.
The total variation distance \cite{Tsybakov2009}
\begin{equation}
  S_1
  = \frac{1}{2} \int \abs*{f_s(x) - f_b(x)} \d x ,
\end{equation}
directly determines the minimal achievable classification error between the two hypotheses and has previously been referred to as the analysis power \cite{Antusch:2024otj}.
The triangular discrimination
\begin{equation} \label{eq:BDT separation}
  S_2
  = \frac{1}{2} \int \frac{\abs*{f_s(x) - f_b(x)}^2}{f_s(x) + f_b(x)} \d x ,
\end{equation}
is called quadratic separation in \software{TMVA} \cite{BaBar:1998yfb, TMVA:2007ngy}.
Unlike $S_1$, this quantity provides a smooth, overlap-weighted measure of the discrepancy between the two probability density functions.
Owing to the quadratic weighting in the numerator, small differences between the densities in sparsely populated regions contribute only weakly to $S_2$.
The quadratic separation is used by \software{TMVA}.

The performance of a \BDT depends on a statistically sound partition of the available data.
The samples used for training, validation, and final evaluation must be mutually independent in order to avoid biased estimates of the analysis sensitivity.
The validation sample is used to monitor and optimise the classifier response, thereby limiting the impact of overfitting to the training data, while the final evaluation is performed only on events not used in either step.
In this work, \qty{10}{\%} of the data is used for training, \qty{10}{\%} for validation, and the remaining \qty{80}{\%} for the final evaluation of the analysis.

A further consideration is whether to train a single \BDT on the full set of signal events or to train separate \BDTs for each point in the parameter scan.
The former approach benefits from a larger combined training sample and yields a unified analysis strategy that can be applied consistently across all benchmark points.
In contrast, the latter approach allows the classifier to exploit features specific to a given mass scale, thereby optimising the discrimination between signal and background locally, without being constrained by performance across the broader parameter space.
The results presented in \cref{sec:results} are produced using this sliding \HNL mass-window technique.

For the background samples, event weights are assigned according to the corresponding production cross sections, so that the training reflects the expected relative background composition.
For classifiers trained on individual benchmark points, the signal does not need to be weighted.

The achieved separation between signal and background is shown as a function of the \HNL mass in \cref{fig:reconstruction:bdt-separation}.
Comparing the two search channels, the separation between signal and background is typically larger in the \LN-blind search, with the notable exception of the \EW scale, where its performance is more strongly degraded than in the \LN-violating search.

The fraction of events in a given category, namely signal or background, that pass a threshold cut, \ie events for which the classifier response exceeds $x$, is given by the survival function of their distribution.
In terms of the probability density function, it reads
\begin{align}
  \epsilon(x) & = 1 - F(x) ,                     &
  F(x)        & = \int_{-\infty}^x f(x') \d x' .
\end{align}
The expected numbers of signal and background events surviving this cut are then determined by the corresponding production cross sections, $\sigma_s$ and $\sigma_b$, the collider luminosity $\mathcal L$, and the threshold $x$ as
\begin{align}
  s(x) & = \mathcal L \sigma_s^{} \epsilon_s(x) ,                                                         &
  b(x) & = \mathcal L \sum_{\scriptstyle\clap{\text{\footnotesize backgrounds}}} \sigma_b \epsilon_b(x) ,
\end{align}
where the signal cross section is scaled to the actual value of the active-sterile mixing parameter \eqref{eq:signal-scaling} and, for the \LN-violating search, to the correct \SS scaling factor \eqref{eq:zeta}.

\subsection{Statistical analysis}\label{sec:statistics}

In a single-bin counting experiment, the likelihood of observing $n$ events when the hypothesis predicts $h$ events is given by the Poisson probability mass function
\begin{equation}
  f_\text{Poisson}(n;h) = \frac{h^n e^{-h}}{n!} .
\end{equation}
The likelihood ratio of a hypothesis $h$ with respect to the saturated hypothesis $h=n$ is then
\begin{equation}
  \lambda(n, h) = \frac{f_\text{Poisson}(n;h)}{f_\text{Poisson}(n;n)} .
\end{equation}
Since the Poisson likelihood is maximised at $h=n$, the likelihood ratio is bounded by
\begin{equation}
  0 \leq \lambda(n, h) \leq 1 ,
\end{equation}
where values close to zero correspond to a hypothesis that describes the data much worse than the saturated hypothesis, while values close to one correspond to a hypothesis that describes the data equally well.
The corresponding likelihood-ratio test statistic is
\begin{equation}
  \frac{q(n, h)}{2} = - \ln \lambda(n, h) = h-n+n\ln\frac nh .
\end{equation}
It becomes large when the hypothesis describes the data poorly and approaches zero when it describes the data as well as the saturated hypothesis.
In the asymptotic approximation for one effective degree of freedom, the corresponding significance in Gaussian standard deviations is
\begin{equation}
  Z(n, h) = \sqrt{q(n, h)} .
  \label{eq:significance}
\end{equation}
To claim exclusion, the signal-plus-background hypothesis $h=b+s$ must be rejected; to claim discovery, the background-only hypothesis $h=b$ must be rejected.
For a known background yield $b$ and signal yield $s$, and in the absence of real data, the median expected significance can be obtained from the Asimov dataset generated under the opposite hypothesis, giving $n=b$ for exclusion and $n=b+s$ for discovery.
Conventionally,
\begin{align}
  Z_\text{discovery} & = Z(b+s,b) \geq \qty{5}{\stddev} , &
  Z_\text{exclusion} & = Z(b,b+s) \geq \qty{2}{\stddev} ,
  \label{eq:significance-exclusion-discovery}
\end{align}
are used as approximate discovery and \qty{95}{\%} confidence-level exclusion criteria, respectively.
For small signal yields, both significances approach the same leading value
\begin{equation}
  Z_\text{exclusion} = Z_\text{discovery} = \frac{s}{\sqrt b} + \order*{\frac{s^2}{b^{3/2}}} .
\end{equation}
These expressions assume that the only uncertainty is the Poisson fluctuation of the event count.
The inclusion of systematic uncertainties in the likelihood would generally reduce the significance.

The threshold $x$ is therefore optimised with respect to the relevant parameters to maximise the significance.
In practice, the optimal threshold is determined by scanning over $x$ and evaluating the significance on the validation dataset at each point.
The value $x^*$ that maximises the significance defines the operating point of the classifier.
This procedure is illustrated in \cref{fig:bdt-distribution}.

\subsection{Reinterpretation of $\LHC$ bounds}

In \SPSSs, observable \LN violation is introduced predominantly by \NNOs and their damping.
These effects are relevant only for \HNLs in $s$-channel processes, \cf \cref{sec:oscillations}.
By comparison, processes mediated by $t$-channel \HNLs are strongly suppressed for small mass splittings.

This observation has important implications for the reinterpretation of existing experimental bounds.
In particular, $t$-channel constraints are generally not applicable in the \SPSS parameter space with small \LN-violating parameters, since the corresponding amplitudes do not benefit from oscillation enhancement and remain negligibly small because of cancellations between mass-eigenstate contributions.
Constraints from $s$-channel processes are more relevant, but they can significantly overestimate the sensitivity to \LN violation if they assume maximally broken \LN, and therefore neglect the suppression that occurs when the mass splitting is much smaller than the decay width and well below the damping threshold.
This effect is illustrated in \cref{fig:lnv-energy-scale}.

To quantify this effect, we consider a benchmark model containing a maximally \LN-violating pseudo-Dirac \HNL with fixed mass and active-sterile mixing $\vec \theta_0$, for which an experimental analysis yields a significance of $Z$.
If the mass splitting is reduced well below the damping threshold, while remaining much smaller than the heavy-neutrino decay width, the effective cross section is suppressed by the corresponding \LN-violating scaling factor in \eqref{eq:zeta}.
As a result, the expected number of signal events decreases, and the original significance can no longer be reached.

To recover the original statistical sensitivity, the active-sterile mixing must be increased until the rate enhancement described by \eqref{eq:signal-scaling} compensates for the suppression from the \LN ratio in \eqref{eq:zeta}
\begin{align}
  \sigma(\vec \theta_0) & = \zeta^{\SS}(R) \sigma(\vec \theta) , &
  R                     & = R(\Gamma_N(\vec \theta), \Delta m) .
\end{align}
In the limit of large decay widths and small mixing angles, the approximations \eqref{eq:zeta limit,eq:signal-scaling-approximation} simplify this equation to
\begin{equation}
  \abs{\vec \theta}^2 = \frac{\abs{\vec \theta_0}^2}{1 - e^{-\lambda}} .
\end{equation}
This expression depends only on the mass splitting and the damping threshold; it highlights how the mixing grows rapidly as the mass splitting approaches zero.

Therefore, existing experimental limits must be reinterpreted by mapping them onto an effective mixing parameter that accounts for the suppression encoded in the \LN ratio.
In practice, exclusion bounds obtained under the assumption of maximal \LN violation should be shifted towards larger mixing angles when applied to \SPSSs with small mass splittings.

In the inverse seesaw regime \eqref{eq:ISS} the \HNL mass splitting is directly tied to the single light-neutrino mass.
Although increasing the active-sterile mixing still enhances the underlying cross section, it also reduces the heavy mass splitting at fixed light-neutrino mass and therefore suppresses observable \LN violation.
For a given light-neutrino mass, this implies an effective upper limit on the active-sterile mixing beyond which the \LN-violating signal becomes increasingly suppressed.

\subsection{Distinguishing pseudo-Dirac and double-Majorana $\HNLs$} \label{sec:distinguishability}

The statistical analysis introduced in \cref{sec:statistics} quantifies the expected significance for discovering or excluding a signal relative to the background-only hypothesis.
Beyond the signal reach itself, a complementary question is whether a pseudo-Dirac pair with suppressed observable \LN violation can be distinguished from the maximally \LN-violating limit.

For fixed mass and active-sterile mixing, we denote the expected \SS signal yield of a pseudo-Dirac pair by $s(R)$ and the corresponding double-Majorana yield by $s(1)$.
Using the likelihood-ratio statistic \eqref{eq:significance}, the expected significances for rejecting either hypothesis are
\begin{align}
  Z_\text{discovery}^\text{pseudo-Dirac} & = Z(b + s(R), b + s(1)) , &
  Z_\text{exclusion}^\text{pseudo-Dirac} & = Z(b + s(1), b + s(R)) .
\end{align}
The first quantity measures the expected power to reject the double-Majorana interpretation if a suppressed pseudo-Dirac rate is realised.
The second measures the expected power to exclude the suppressed pseudo-Dirac interpretation under the maximally \LN-violating rate hypothesis.

\begin{figure}
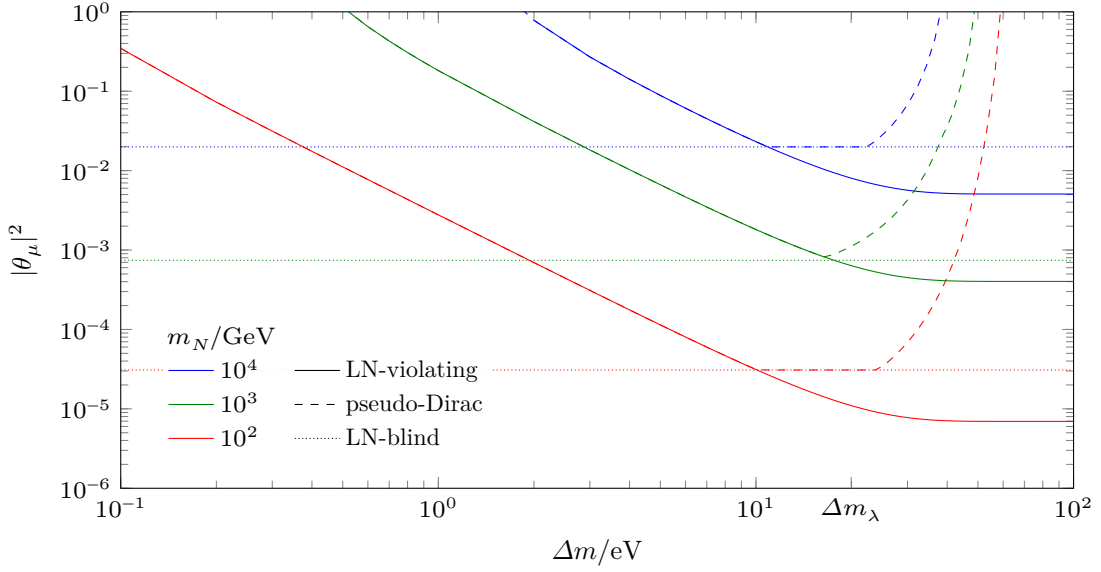

  \includepgf{deltam-mixing-z2}
  \caption[Distinguishing pseudo-Dirac and double-Majorana \HNLslong]{
    Expected sensitivity to distinguishing a pseudo-Dirac \HNL pair with suppressed observable \LN violation from the double-Majorana limit at the \FCChh.
    The displayed quantity combines the \LN-blind and \LN-violating exclusion sensitivities with the power to exclude the suppressed pseudo-Dirac hypothesis \eqref{eq:distinguishability}.
  }
  \label{fig:deltam-mixing-z2}
\end{figure}

A \SS rate alone is not sufficient to determine the \LN ratio.
The \LN-violating yield depends both on the active-sterile mixing and on the suppression factor \eqref{eq:zeta}; a smaller mixing in the double-Majorana limit can therefore mimic a larger mixing with a reduced \LN ratio.
This degeneracy is broken only if the signal normalisation is constrained independently, for example by a \LN-conserving or -blind channel.

We therefore require three ingredients for the exclusion-based distinguishability criterion used here:
\begin{inlinelist}
  \item sufficient \LN-blind exclusion reach for the signal normalisation
  \item sufficient \LN-violating exclusion reach for the \SS signal
  \item sufficient model-comparison power to exclude the suppressed pseudo-Dirac hypothesis
\end{inlinelist}
In the present work we quantify this by
\begin{equation} \label{eq:distinguishability}
  Z_{\min} = \min\left( Z_\text{exclusion}^\text{\LN-blind}, Z_\text{exclusion}^\text{\LN-violating}, Z_\text{exclusion}^\text{pseudo-Dirac} \right) .
\end{equation}
The use of the minimum enforces the three requirements simultaneously: both search channels must have sufficient exclusion sensitivity, and the double-Majorana \SS rate must be sufficiently separated from the suppressed pseudo-Dirac rate at the same mass and mixing.

When the mass splitting is much smaller than both the heavy-neutrino decay width and the damping threshold, the \LN ratio is close to zero and the \LN-violating channel lacks the required sensitivity.
When the \LN ratio approaches one, either because of \NNOs or their damping, the pseudo-Dirac and double-Majorana hypotheses become experimentally indistinguishable in rate-based observables, even though \LN violation itself may remain discoverable.
Only the intermediate regime can therefore satisfy all three contributions to the significance \eqref{eq:distinguishability}.
This dependence is illustrated in \cref{fig:deltam-mixing-z2}.

\section{Results} \label{sec:results}

In the following, we present the results for \LN-blind and \LN-violating searches for pseudo-Dirac \HNLs on the mass-coupling plane.
We interpret both the analysis and the \LHC bounds under the assumption that the active-sterile mixing occurs exclusively with second-generation leptons.
All predicted \FCChh exclusion limits are given at the \qty{95}{\%} confidence level, corresponding to a significance of \qty{2}{\stddev}.
All projected sensitivities are obtained using the sliding \HNL mass-window procedure described in \cref{sec:BDT}.

The projected sensitivities are compared to existing constraints from \LHC searches.
Throughout this section, shaded grey and rainbow-coloured regions indicate parameter space covered by $s$-channel \LHC searches.
These searches fall into three categories:
\begin{inlinelist}
  \item regions that are excluded by \LN-blind searches are shown in dark grey
  \item regions that are covered by \LN-violating searches are either shown in grey or reinterpreted as mass-splitting-dependent bounds shown in rainbow colours
  \item regions that are covered by searches that conflate \LN violation with \LF violation are shown in light grey or reinterpreted as mass-splitting-dependent bounds shown in light rainbow colours
\end{inlinelist}
Dotted regions indicate bounds from \LN-violating searches with \HNL exchange in the $t$-channel, which do not apply in the symmetry-protected regime.
These different categories of \LHC bounds are discussed in \cref{sec:LHC}.
Hashed regions denote the non-perturbative regime in which the \HNL Yukawa coupling \eqref{eq:yukawa coupling} satisfies $y_1 \geq 8\pi$.

The sensitivity of \LN-blind searches is approximately independent of the amount of observable \LN violation and therefore probes the maximal collider reach for \HNL production.
In contrast, the sensitivity of \LN-violating searches depends on the \LN ratio \eqref{eq:LN ratio}, which is controlled by the interplay of \NNOs and decoherence effects.
Consequently, the interpretation of \LN-violating searches differs significantly across the small- and large-\LN-ratio parts of the oscillation- and damping-enhanced regimes introduced in \cref{sec:regimes}.

\begin{figure}
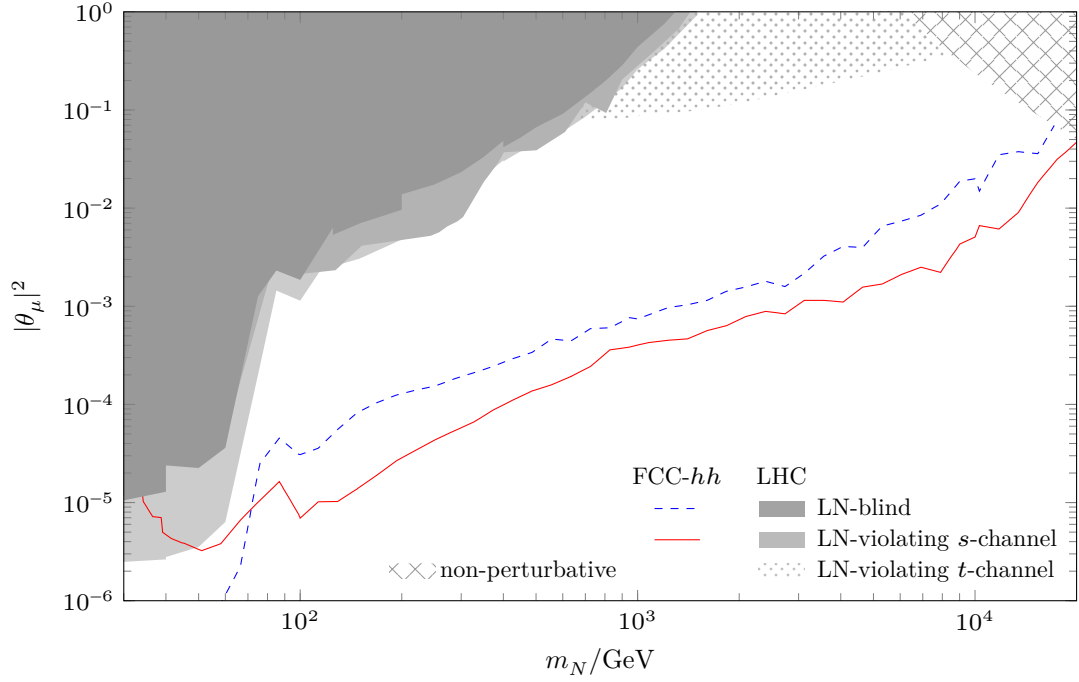

  \includepgf{result}
  \caption[Maximal sensitivities]{
    Expected sensitivities on the mass-coupling plane for pseudo-Dirac \HNLs with maximal observable \LN violation at the \FCChh.
    The projected sensitivities of the \LN-blind and \LN-violating searches are shown using a sliding \HNL mass window.
    These limits correspond to the maximal exclusion reach attainable in pseudo-Dirac \HNL scenarios.
    The dark grey region denotes existing bounds from \LN-blind searches at the \LHC.
    The grey regions indicate limits from \LN-violating $s$-channel searches interpreted in the double-Majorana limit.
    The light grey regions indicate limits from \LN-violating $s$-channel searches that conflate \LN violation and \LF violation, interpreted in the double-Majorana limit.
    The dotted region shows limits from \LN-violating searches with \HNL exchange in the $t$-channel, which are not applicable in the symmetry-protected regime.
    See \cref{sec:LHC} for a discussion of the existing bounds.
    The hashed region corresponds to the non-perturbative regime.
  } \label{fig:exclusion limits}
\end{figure}

The maximal exclusion reach for pseudo-Dirac \HNLs with maximal observable \LN violation is shown in \cref{fig:exclusion limits}.
Over most of the considered mass range, the search strategy targeting \LN violation discussed in \cref{sec:LN violation} significantly outperforms the \LN-blind search described in \cref{sec:LN blind}.
This improvement is driven by the substantially smaller \SM backgrounds in the \SS dimuon channel.
As a consequence, the \LN-violating search extends the exclusion reach in the active-sterile mixing by up to approximately one order of magnitude compared to the \LN-blind search.

For \HNL masses below approximately the $W$ boson mass, the \LN-blind search is more sensitive than the \LN-violating search.
This result follows from the smaller cross section, reconstruction efficiency, and signal-background separation of the latter analysis, all of which are depicted in \cref{fig:reconstruction}.

\begin{figure}
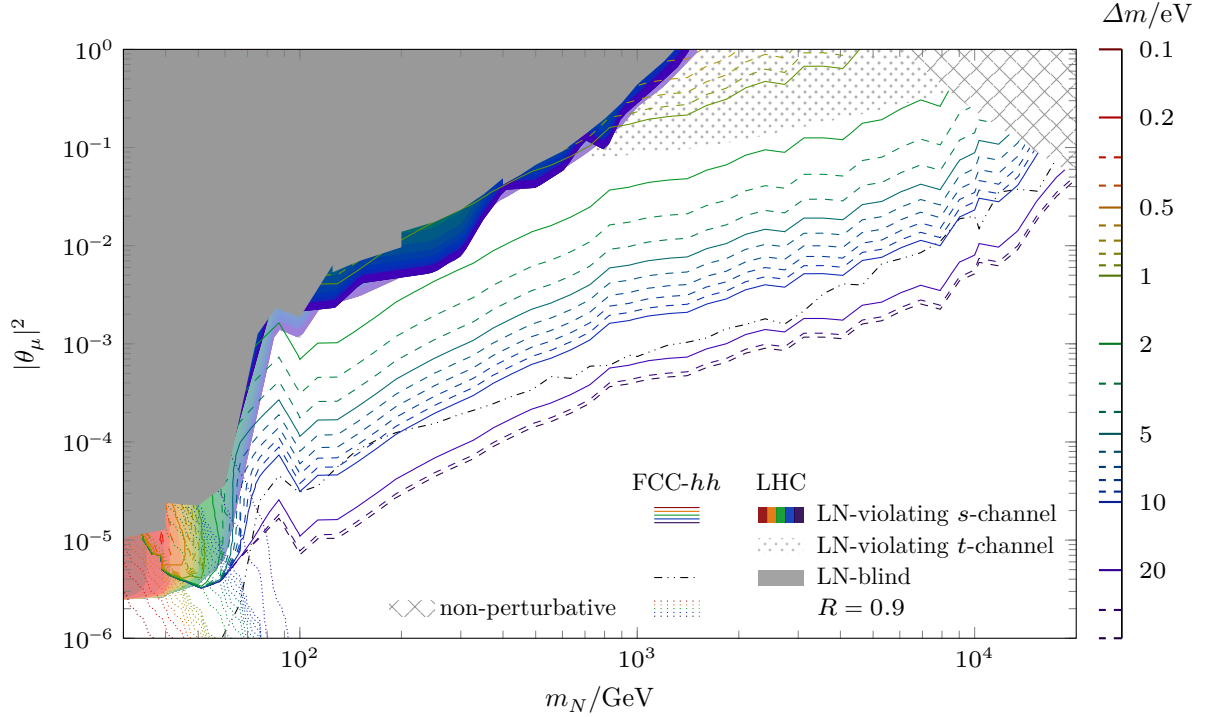

  \includepgf{result-promptlnv-deltam}
  \caption[Sensitivity to \LNlong violation]{
    Expected sensitivity from searches for \LN violation as a function of the pseudo-Dirac \HNL mass splitting.
    The strongest sensitivity is obtained in the large-\LN-ratio regime, where either oscillations or damping drive observable \LN violation close to its maximal value.
    In the small-\LN-ratio regimes, the suppression of the \LN ratio leads to a correspondingly weaker sensitivity.
    The double-dotted-dashed line indicates the reach of the \LN-blind search.
    The dotted lines indicate a \LN ratio of $0.9$ and mark the transition towards the double-Majorana limit.
    The rainbow-coloured regions show the reinterpretation of existing bounds from resonant \LN-violating $s$-channel searches for different pseudo-Dirac mass splittings.
    The light rainbow-coloured regions show the reinterpretation of existing bounds from searches that conflate \LN violation with \LF violation for different pseudo-Dirac mass splittings.
    The grey region denotes existing bounds from \LN-blind searches at the \LHC.
    The dotted region indicates bounds from \LN-violating searches with \HNL exchange in the $t$-channel, which are not applicable in the symmetry-protected regime.
    See \cref{sec:LHC} for a discussion of the existing bounds.
    The hashed region corresponds to the non-perturbative regime.
  } \label{fig:LNV exclusion limits}
\end{figure}

The projected \FCChh sensitivities extend well beyond existing \LHC constraints across the entire mass range considered.
The limits shown in \cref{fig:exclusion limits} therefore represent the maximal exclusion reach attainable for pseudo-Dirac \HNLs.
The dependence of the sensitivity on the pseudo-Dirac mass splitting is investigated in \cref{fig:LNV exclusion limits}.

In \SPSSs, observable \LN violation is generically suppressed relative to the double-Majorana limit commonly assumed when interpreting searches for \LN violation in terms of a single Majorana \HNL.
As discussed in \cref{sec:SPSS}, the amount of observable \LN violation is controlled by the mass splitting of the pseudo-Dirac pair.
The resulting sensitivity is shown in \cref{fig:LNV exclusion limits}.

When the \LN ratio approaches unity, either because oscillations dominate at small decay widths or because damping dominates for mass splittings sufficiently above the damping threshold, the exclusion reach approaches the maximal reach shown in \cref{fig:exclusion limits}.
When the \LN ratio is small, the sensitivity of the \LN-violating search is correspondingly reduced.
Nevertheless, in the region where the present analysis improves upon existing searches, mass splittings down to a few tenths of an eV remain accessible.
For mass splittings above approximately \qty{30}{eV}, the pseudo-Dirac pair becomes phenomenologically indistinguishable from the double-Majorana limit.
In the relevant part of the parameter space, this transition occurs in the large-\LN-ratio damping-enhanced regime discussed in \cref{sec:regimes} and is consistent with the behaviour of the \LN ratio shown in \cref{fig:Rll as function of mass}.

The minimal linear seesaw provides two representative benchmarks.
For \NO and \IO, it predicts mass splittings of \qty{41.4}{meV} and \qty{762}{\micro eV}, respectively; see \cref{tab:neutrino-masses} and \cite{Antusch:2023nqd}.
Therefore, these benchmarks are close to the Dirac limit, leading to a strongly suppressed \LN-violating signal and a correspondingly reduced sensitivity.

\begin{figure}
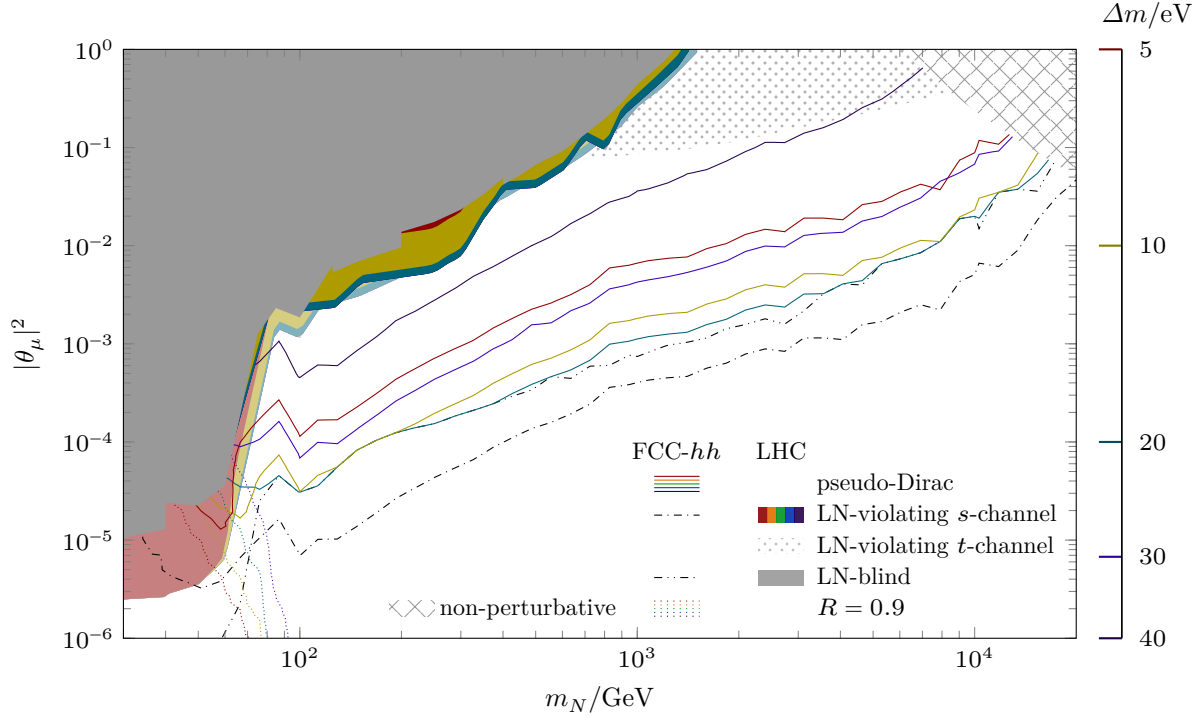

  \includepgf{result-promptlnv-deltam-2}
  \caption[Distinguishability of pseudo-Dirac and double-Majorana $\HNLs$]{
    Expected sensitivity to distinguishing a pseudo-Dirac \HNL from the double-Majorana limit according to the criterion defined in \eqref{eq:distinguishability}.
    The solid lines indicate where the \LN-blind and \LN-violating exclusion sensitivities are combined with the exclusion of the corresponding pseudo-Dirac rate hypothesis.
    The dotted-dashed line reproduces the maximal \LN-violating sensitivity shown in \cref{fig:exclusion limits} and therefore represents the maximal parameter space in which a distinguishability test can be performed.
    The double-dotted-dashed line indicates the reach of the \LN-blind search.
    The dotted lines indicate a \LN ratio of $0.9$ and mark the transition towards the double-Majorana limit.
    The rainbow-coloured regions show the reinterpretation of existing limits from resonant \LN-violating $s$-channel searches for pseudo-Dirac \HNLs with different mass splittings.
    The light rainbow-coloured regions show the reinterpretation of existing searches that conflate \LN violation with \LF violation for pseudo-Dirac \HNLs with different mass splittings.
    The grey region denotes existing bounds from \LN-blind searches at the \LHC.
    The dotted region indicates bounds from \LN-violating searches with \HNL exchange in the $t$-channel, which are not applicable in the symmetry-protected regime.
    See \cref{sec:LHC} for a discussion of the existing bounds.
    The hashed region corresponds to the non-perturbative regime.
  } \label{fig:distinguishability}
\end{figure}

The reach for distinguishing a pseudo-Dirac \HNL from the double-Majorana limit is shown in \cref{fig:distinguishability}.
Compared to the \LN-violating exclusion reach shown in \cref{fig:exclusion limits}, the distinguishability reach is restricted to a smaller region of parameter space.
As discussed in \cref{sec:distinguishability}, the criterion requires sufficient exclusion sensitivity in both the \LN-blind and \LN-violating channels, as well as a significant separation between the pseudo-Dirac and double-Majorana \SS rates.
These requirements are combined in the distinguishability criterion \eqref{eq:distinguishability}.

For very small mass splittings, the \LN-violating signal is strongly suppressed and does not provide sufficient sensitivity.
Conversely, for large mass splittings, the \LN ratio approaches unity and the pseudo-Dirac pair becomes phenomenologically indistinguishable from the double-Majorana limit.
Consequently, the distinguishability criterion can only be satisfied for intermediate mass splittings, where \LN violation provides sufficient \SS sensitivity while the pseudo-Dirac rate remains sufficiently different from the double-Majorana expectation.
This behaviour reflects the non-trivial interplay between \NNOs and decoherence effects that governs the amount of observable \LN violation in pseudo-Dirac \HNL scenarios.

\begin{figure}
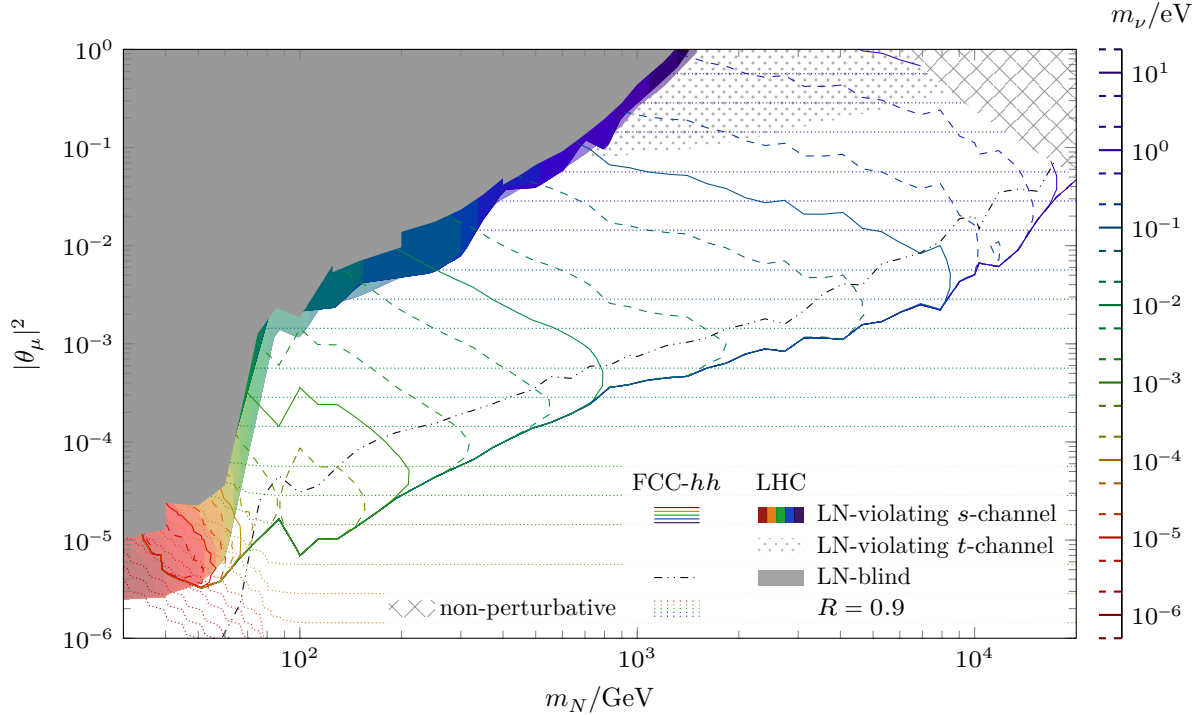

  \includepgf{result-promptlnv-mv}
  \caption[Sensitivity to \LNlong violation in the inverse seesaw]{
    Expected sensitivity from searches for \LN violation in the inverse seesaw as a function of the light neutrino mass.
    The displayed lines show the sensitivity obtained by relating the pseudo-Dirac mass splitting to the light neutrino mass through \eqref{eq:inverse seesaw mass splittings}.
    The dotted lines indicate a \LN ratio of $0.9$ and mark the transition towards the double-Majorana limit.
    The double-dotted-dashed line indicates the reach of the \LN-blind search.
    The rainbow-coloured regions show the reinterpretation of existing limits from resonant \LN-violating $s$-channel searches using the inverse-seesaw prediction for the pseudo-Dirac mass splitting.
    The light rainbow-coloured regions show the reinterpretation of existing searches that conflate \LN violation with \LF violation using the inverse-seesaw prediction for the pseudo-Dirac mass splitting.
    The grey region denotes existing bounds from \LN-blind searches at the \LHC.
    The dotted region indicates bounds from \LN-violating searches with \HNL exchange in the $t$-channel, which are not applicable in the symmetry-protected regime.
    See \cref{sec:LHC} for a discussion of the existing bounds.
    The hashed region corresponds to the non-perturbative regime.
  } \label{fig:exclusion limits inverse seesaw}
\end{figure}

The results can be reinterpreted within the inverse seesaw framework \eqref{eq:ISS}.
In this scenario, the pseudo-Dirac mass splitting is directly related to the light neutrino mass through \eqref{eq:inverse seesaw mass splittings}.
The corresponding sensitivity is shown in \cref{fig:exclusion limits inverse seesaw}.

For small neutrino masses, the pseudo-Dirac mass splitting is strongly suppressed, resulting in a correspondingly small amount of observable \LN violation.
As the neutrino mass increases, the larger mass splitting enhances the \LN ratio and improves the sensitivity of the \LN-violating search.
Consequently, collider searches for \LN violation become sensitive to light neutrino masses down to approximately \qty{0.1}{meV}.

The inverse seesaw with a single pseudo-Dirac \HNL generates only one non-vanishing light neutrino mass and therefore cannot account for the observed pattern of neutrino flavour oscillations.
The light neutrino mass shown in \cref{fig:exclusion limits inverse seesaw} should therefore be interpreted as an effective parameter controlling the pseudo-Dirac mass splitting, rather than as a realistic neutrino-mass spectrum.
For this reason, cosmological constraints on the sum of light neutrino masses have not been included.

\begin{figure}
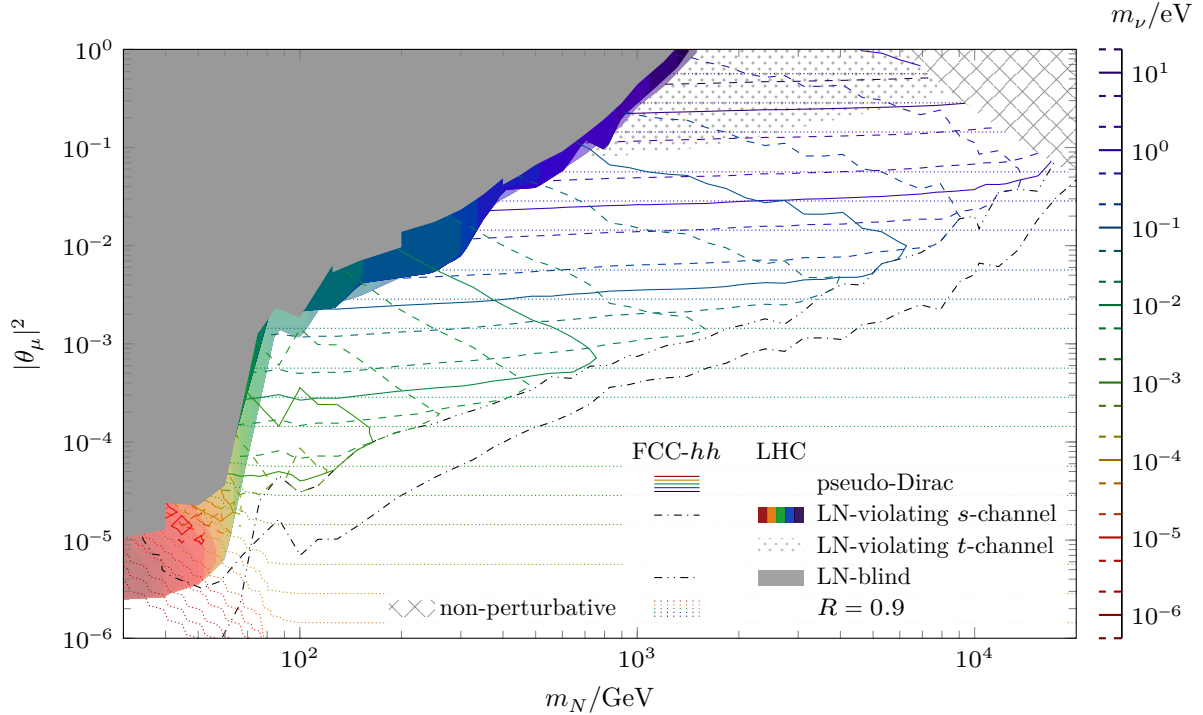

  \includepgf{result-promptlnv-mv-2}
  \caption[Distinguishability of pseudo-Dirac and double-Majorana $\HNLs$ in the inverse seesaw]{
    Expected sensitivity to distinguishing a pseudo-Dirac \HNL from the double-Majorana limit in the inverse seesaw as a function of the light neutrino mass.
    The displayed lines indicate where the \LN-blind and \LN-violating exclusion sensitivities are combined with the exclusion of the corresponding pseudo-Dirac rate hypothesis.
    The dotted-dashed line reproduces the maximal \LN-violating sensitivity shown in \cref{fig:exclusion limits inverse seesaw} and therefore represents the maximal parameter space in which a distinguishability test can be performed.
    The double-dotted-dashed line indicates the reach of the \LN-blind search.
    The dotted lines indicate a \LN ratio of $0.9$ and mark the transition towards the double-Majorana limit.
    The rainbow-coloured regions show the reinterpretation of existing limits from resonant \LN-violating $s$-channel searches using the inverse-seesaw prediction for the pseudo-Dirac mass splitting.
    The light rainbow-coloured regions show the reinterpretation of searches that conflate \LN violation with \LF violation using the inverse-seesaw prediction for the pseudo-Dirac mass splitting.
    The grey region denotes existing bounds from \LN-blind searches at the \LHC.
    The dotted region indicates bounds from \LN-violating searches with \HNL exchange in the $t$-channel, which are not applicable in the symmetry-protected regime.
    See \cref{sec:LHC} for a discussion of the existing bounds.
    The hashed region corresponds to the non-perturbative regime.
  } \label{fig:distinguishability inverse seesaw}
\end{figure}

The reach for distinguishing a pseudo-Dirac HNL from the double-Majorana limit within the inverse seesaw is shown in \cref{fig:distinguishability inverse seesaw}.
Compared to the \LN-violating exclusion reach shown in \cref{fig:exclusion limits inverse seesaw}, the parameter space in which a pseudo-Dirac \HNL can be distinguished from the double-Majorana limit is significantly reduced.
As discussed in \cref{sec:distinguishability}, satisfying the criterion requires sufficient exclusion sensitivity in both search channels and a statistically significant separation between the pseudo-Dirac and double-Majorana \SS rates.

For small neutrino masses, the pseudo-Dirac mass splitting provides sufficient \LN-violating sensitivity only at small mixing angles.
Conversely, for large neutrino masses, large mixing angles are needed for the \LN ratio to differ appreciably from unity.
Therefore, for each neutrino mass, the distinguishability criterion can be satisfied only within a limited range of the active-sterile mixing angle, in which \LN violation provides sufficient \SS sensitivity while the pseudo-Dirac and double-Majorana rates remain experimentally separable.
This behaviour mirrors the interplay between \NNOs and decoherence effects discussed in \cref{sec:regimes} and demonstrates that the \FCChh can probe a substantial region of inverse-seesaw parameter space in which the pseudo-Dirac and double-Majorana rate hypotheses can be separated experimentally.

\section{Conclusion} \label{sec:conclusion}

The observation of \HNLs would provide direct evidence for physics beyond the \SM and could shed light on the origin of neutrino masses.
In type~I seesaw models with \HNL masses near the \EW scale, the smallness of the light neutrino masses must be protected by a symmetry in order to avoid fine-tuning.
In \SPSSs, this role is played by an approximately conserved \LN-like symmetry,
whose small breaking generates the observed neutrino masses.
As a consequence, the heavy neutrinos form pseudo-Dirac pairs with small mass splittings.

We have investigated the impact of these mass splittings on collider searches for \LN violation.
While \LN violation is suppressed at the amplitude level in plane-wave \QFT by the small symmetry-breaking parameters, \LN violation can nevertheless be observable due to \NNOs and decoherence effects.
The resulting phenomenology is governed by oscillation- and damping-enhanced regimes, each of which can yield either a small or large \LN ratio.
In the small-\LN-ratio regimes, observable \LN violation is enhanced relative to the plane-wave expectation, although the resulting signal remains strongly suppressed.
In the large-\LN-ratio regimes, oscillations or damping can render the pseudo-Dirac pair effectively equivalent to two independent Majorana neutrinos in rate-based observables, yielding maximal observable \LN violation.
Observable \LN violation can therefore drastically exceed the amount expected from the underlying Lagrangian in plane-wave \QFT.

To explore the collider implications of this behaviour, we performed a study of \LN-violating \SS dimuon signatures at the \FCChh and compared the results to a corresponding \LN-blind search.
Assuming maximal observable \LN violation, the \LN-violating search improves the sensitivity to the active-sterile mixing by up to approximately one order of magnitude because of the substantially reduced \SM backgrounds.
However, we found that the sensitivity depends strongly on the pseudo-Dirac mass splitting.
In particular, the reach of \LN-violating searches can be significantly reduced in the small-\LN-ratio regimes and approaches the conventional double-Majorana expectation only in the large-\LN-ratio regimes.
This observation demonstrates that collider limits from \LN-violating searches cannot, in general, be interpreted independently of the pseudo-Dirac mass splitting.

We further investigated the extent to which a pseudo-Dirac \HNL can be distinguished experimentally from the double-Majorana limit.
In the exclusion-based criterion used here, this requires \LN-blind and \LN-violating exclusion sensitivity together with a pseudo-Dirac exclusion term large enough to separate the pseudo-Dirac and double-Majorana rate hypotheses.
We found that such a distinction is possible for intermediate mass splittings, where the \LN ratio provides sufficient \LN-violating sensitivity while remaining sufficiently below unity for the two rate hypotheses to be separated.

Finally, we interpreted our results within the inverse seesaw framework.
In this scenario, the pseudo-Dirac mass splitting is directly related to the single non-zero light neutrino mass.
We showed that a future hadron collider operating at \FCChh energies can probe \LN-violating \HNL processes over a substantial region of inverse-seesaw parameter space and, in favourable scenarios, separate the pseudo-Dirac and double-Majorana rate hypotheses experimentally.

Our results highlight that searches for \LN violation at colliders probe not only the presence of \LN-violating interactions, but also the quantum dynamics of pseudo-Dirac neutrinos via the \LN ratio.
A consistent interpretation of present and future collider searches for \HNLs therefore requires taking the interplay of symmetry protection, \NNOs, and decoherence effects into account.

\subsection*{Acknowledgements}

The work of Bruno M.\ S.\ Oliveira is supported by the PhD Studentship \textnumero\ 2024.\allowbreak 00923.\allowbreak BD from the Portuguese \FCT.
The work of Jan Hajer is supported by the \FCT project \textnumero\ \href{https://doi.org/10.54499/2020.03969.CEECIND/CP1587/CT0014}{2020.\allowbreak 03969.\allowbreak CEECIND/\allowbreak CP1587/\allowbreak CT0014}.
The work of Bruno M.\ S.\ Oliveira is supported by the \FCT project \href{https://doi.org/10.54499/UID/00777/2025}{UID/\allowbreak 00777/\allowbreak 2025}.
The work of Bruno M.\ S.\ Oliveira and Stefan Antusch was supported by the \CHEF, with funding provided specifically by \SERI and by the University of Basel.
Bruno M.\ S.\ Oliveira and Jan Hajer thank the Particles and Cosmology group at the University of Basel for their hospitality.

\end{document}